\documentclass[twocolumn, 11pt]{aastex63}%linenumbers
\usepackage{epsfig,xcolor}
\usepackage{natbib}
\usepackage[normalem]{ulem}
\usepackage{comment}
\usepackage{appendix}
\usepackage{mwe}

\begin{document}

\title{A Multi-Wavelength Technique for Estimating Galaxy Cluster Mass Accretion Rates}

\author{John Soltis}
\affiliation{Department of Physics \& Astronomy, Johns Hopkins University, Baltimore, MD 21218, USA}
\author{Michelle Ntampaka}
\affiliation{Space Telescope Science Institute, Baltimore, MD 21218, USA}
\affiliation{Department of Physics \& Astronomy, Johns Hopkins University, Baltimore, MD 21218, USA}

\author{Benedikt Diemer}
\affiliation{Department of Astronomy, University of Maryland, College Park, MD 20742, USA}
\author{John ZuHone}
\affiliation{Center for Astrophysics $\mid$ Harvard \& Smithsonian, 60 Garden Street, Cambridge, MA 02138, USA}

\author{Sownak Bose}
\affiliation{Institute for Computational Cosmology, Department of Physics,
University of Durham, South Road, Durham, DH1 3LE, UK}

\author{Ana Maria Delgado}
\affiliation{Department of Physics \& Astronomy, Johns Hopkins University, Baltimore, MD 21218, USA}

\author{Boryana Hadzhiyska}
\affiliation{Miller Institute for Basic Research in Science, University of California, Berkeley, CA, 94720, USA}
\affiliation{Physics Division, Lawrence Berkeley National Laboratory, Berkeley, CA 94720, USA}
\affiliation{Berkeley Center for Cosmological Physics, Department of Physics, University of California, Berkeley, CA 94720, USA}

\author{C\'esar Hern\'andez-Aguayo}
\affiliation{Max-Planck-Institut f\"ur Astrophysik, Karl-Schwarzschild-Str. 1,
D-85748, Garching, Germany\\%
Excellence Cluster ORIGINS, Boltzmannstrasse 2, D-85748 Garching,
Germany}

\author{Daisuke Nagai}
\affiliation{Department of Physics, Yale University, New Haven, CT 06520, USA}

\author{Hy Trac}
\affiliation{Department of Physics, Carnegie Mellon University, Pittsburgh, PA 15213, USA}
\affiliation{McWilliams Center for Cosmology and Astrophysics, Carnegie Mellon University, Pittsburgh, PA 15213, USA}

\begin{abstract}
    The mass accretion rate of galaxy clusters is a key factor in determining their structure, but a reliable observational tracer has yet to be established. We present a state-of-the-art machine learning model for constraining the mass accretion rate of galaxy clusters from only X-ray and thermal Sunyaev-Zeldovich observations. Using idealized mock observations of galaxy clusters from the MillenniumTNG simulation, we train a machine learning model to estimate the mass accretion rate. The model constrains 68\% of the mass accretion rates of the clusters in our dataset to within 33\% of the true value without significant bias, a $\sim58$\% reduction in the scatter over existing constraints. We demonstrate that the model uses information from both radial surface brightness density profiles and asymmetries.
\end{abstract}

\section{Introduction}\label{introduction}

Galaxy clusters, the most massive gravitationally bound objects in the Universe, contain collections of tens to thousands of galaxies. Galaxy clusters are roughly 80\% dark matter and 20\% ordinary matter. Of the ordinary matter, the vast majority is a hot, ionized gas known as the intra-cluster medium (ICM). Nearer to the cluster core, this ionized gas is X-ray bright \citep[e.g.,][]{Bulbul_2024}, emitting X-rays via bremsstrahlung, collisional excitation, recombination radiation, and 2-photon emission processes. The ICM is also observable from the inverse-Compton scattering of cosmic microwave background photons off the electrons in the ICM \citep[e.g.,][]{Hilton_2021}, an effect known as the thermal Sunyaev-Zeldovich (tSZ) effect \citep{SZ_1972}. As high density peaks of the mass distribution of the Universe, the abundance of galaxy clusters is sensitive to changes in cosmology \citep[e.g.,][]{eROSITA_Cluster_Abundances_2024}. Galaxy clusters are also an effective probe dark matter physics \citep[e.g.,][]{Andrade_2022} and astrophysics \citep[e.g.,][]{Fournier_2024}.

Galaxy clusters form by accreting dark matter, gas, galaxies, galaxy groups, and smaller galaxy clusters in their local environment \citep[for a review, see][]{Kravtsov_2012}. Simulations suggest that the formation history of a galaxy cluster influences its morphology substantially, including its concentration and shape of its density profile, its ellipticity, and the fraction of its mass contained within its substructure \citep[e.g.,][]{Wong_2012, Diemer_2014, Jiang_2016, Chen_2019, Lau_2021, Mendoza_2023}. Galaxy cluster mass estimates, needed for cosmological constraints, are often dependent on mass proxies, such as the X-ray gas temperature. These proxies can be biased by the physical disturbances induced by mass accretion \citep[e.g.,][]{Lau_2015, Lee_2023, Zhuravleva_2023}. The intensity of disruption and physical disturbance in a galaxy cluster, a direct result of its mass accretion history, is known as the dynamical state. For a review of the biases in galaxy cluster mass counts, see \citet{Pratt_2019}.

Given the importance of galaxy cluster mass accretion rates (MARs), much work has been done to develop proxies, constraints, and estimates of its significance. Correlations have been found between morphology and MAR (or dynamical state) \citep[e.g.,][]{Gouin_2021} and efforts have been undertaken to construct MAR or dynamical state proxies from morphological parameters calculated from observations \citep[e.g.,][]{Perez_2023}. More direct estimates include using constituent galaxies' spectroscopy \citep[e.g.,][]{Pizzardo_2023}, using a classifying machine learning (ML) model to identify clusters with relaxed or merging dynamical state \citep{Arendt_2024}, or connecting mass accretion rates directly to halo and galaxy properties \citep{Mendoza_2023}. 

A direct constraint of galaxy cluster MARs has many potential benefits. Since X-ray or SZ mass estimation bias is correlated with MAR \citep{Nelson2012,Nelson2014a,Nelson2014b}, knowing the MAR would allow one to correct the mass estimation bias more effectively \citep{Shi2016}, thereby improving the cosmological constraints. Alternatively, ML mass estimators might use MAR information to produce more accurate constraints \citep[e.g.,][]{Ho_2023}. It would allow for a more direct disentanglement of the role of MAR on divergent galaxy cluster core behavior, on the influence that MAR has on the shape of the radial density profile, and on the stellar productivity of cluster galaxies. Other interesting astrophysical phenomena, like the formation history of clusters, the role of cluster environment, and the influence of formation history on active galactic nuclei (AGN) activity could be better probed with direct estimates of galaxy cluster MARs. Moreover, the global distribution of galaxy cluster MARs is itself sensitive to cosmology and could offer a new probe to constrain the amplitude of the variance of the matter density distribution \citep[][]{Amoura_2024}.

In this work, we present a method for estimating the MAR of galaxy clusters directly from X-ray and tSZ observations. We choose X-ray and tSZ specifically because they probe the dynamics of the ICM, and are an important means of observing clusters \citep[e.g.,][]{Hilton_2021, SPT_catalog, Bulbul_2024} and constraining cosmology with mass abundances \citep{SPT_Abundances, eROSITA_Cluster_Abundances_2024}. We train a ML model on mock observations of simulated clusters from the MillenniumTNG (MTNG) simulation \citep{MTNG_1, MTNG_2, MTNG_3, MTNG_4, MTNG_5, MTNG_6, MTNG_7, MTNG_8, MTNG_9, MTNG_10} in order to predict the change in $\log M_{200m}$ over one dynamical time (defined in Section~\ref{data}). 

Our model, utilizing a form of neural network known as a  ``Normalizing Flow'' \citep{Papamakarios_2019}, provides an accurate estimate of the probability distribution of potential MARs for a given galaxy cluster. This work builds on a well-established field of applying ML to the study of galaxy clusters. In addition to the aforementioned \citet{Arendt_2024}, deep learning models have been applied to the study of galaxy clusters to constrain mass \citep{Ntampaka2019, Ho_2023, Krippendorf_2024}, improve observations \citep{Soltis_2022}, constrain cosmology \citep{Qiu_2024}, characterize substructure \citep{Henriksen_2024}, classify the membership of galaxy clusters \citep{Farid2023}, identify nearby filaments \citep{Weaver_2023}, and emulate hydrodynamic cosmological simulation observations \citep{Rothschild_2022}. Deep learning and, more specifically, normalizing flows have also been applied to a variety of astronomical effects to great effect \citep[e.g.,][]{Sweere_2022}.

The paper is organized as follows. In Section~\ref{methods}, we discuss the data we used to train the ML model (\S\ref{data}), the ML model architecture (\S\ref{models}), the non-ML model mass accretion rate fitting function (\S\ref{Diemer_2020}), and the training of the ML model (\S\ref{training}). In Section~\ref{results}, we present the results of our work, including the accuracy of the model (\S\ref{accuracy}), the uncertainty estimates of the model (\S\ref{uncertainty}), analysis of the biases of the model (\S\ref{biases}), and the impact of data and model variants (\S\ref{data_robustness}). In Section~\ref{interpretation}, we describe an interpretation method that we used to better understand the model, decomposing images into radially symmetric and asymmetric components. Finally, we review the results, discuss their implications, and explore future potential work in Section~\ref{conclusion}.

\section{Methodology}\label{methods}
Our choice of method and data are constrained by the problem at hand. We expect the ICM, which is the most collisional component of the galaxy cluster, to be the most visibly affected by mass accretion. Therefore, X-ray and tSZ observations are natural choices, because both directly probe the density and temperature of the ICM. Furthermore, more clusters have been identified in the X-ray and tSZ than by other means. Using the ICM as a probe of MARs means that we are sensitive to astrophysics, so we need a hydrodynamic simulation. ML models, especially convolutional neural networks, are excellent at mapping complicated nonlinear relationships like what we expect to exist between observations and MARs. To train a ML model, we need a large, varied, and accurate dataset. Moreover, we want the distribution of MARs to accurately represent the true distribution of MARs in the Universe, which is best achieved using a cosmological simulation. From these conditions, we know that we need to use a convolutional neural network on mock X-ray and tSZ observations of simulated galaxy clusters from a very large, state-of-the-art hydrodynamic cosmological simulation.

\subsection{Data}\label{data}
\begin{figure}%[b]
    \centering
    \null \vspace{-5pt}
    \includegraphics[width=3.2in]{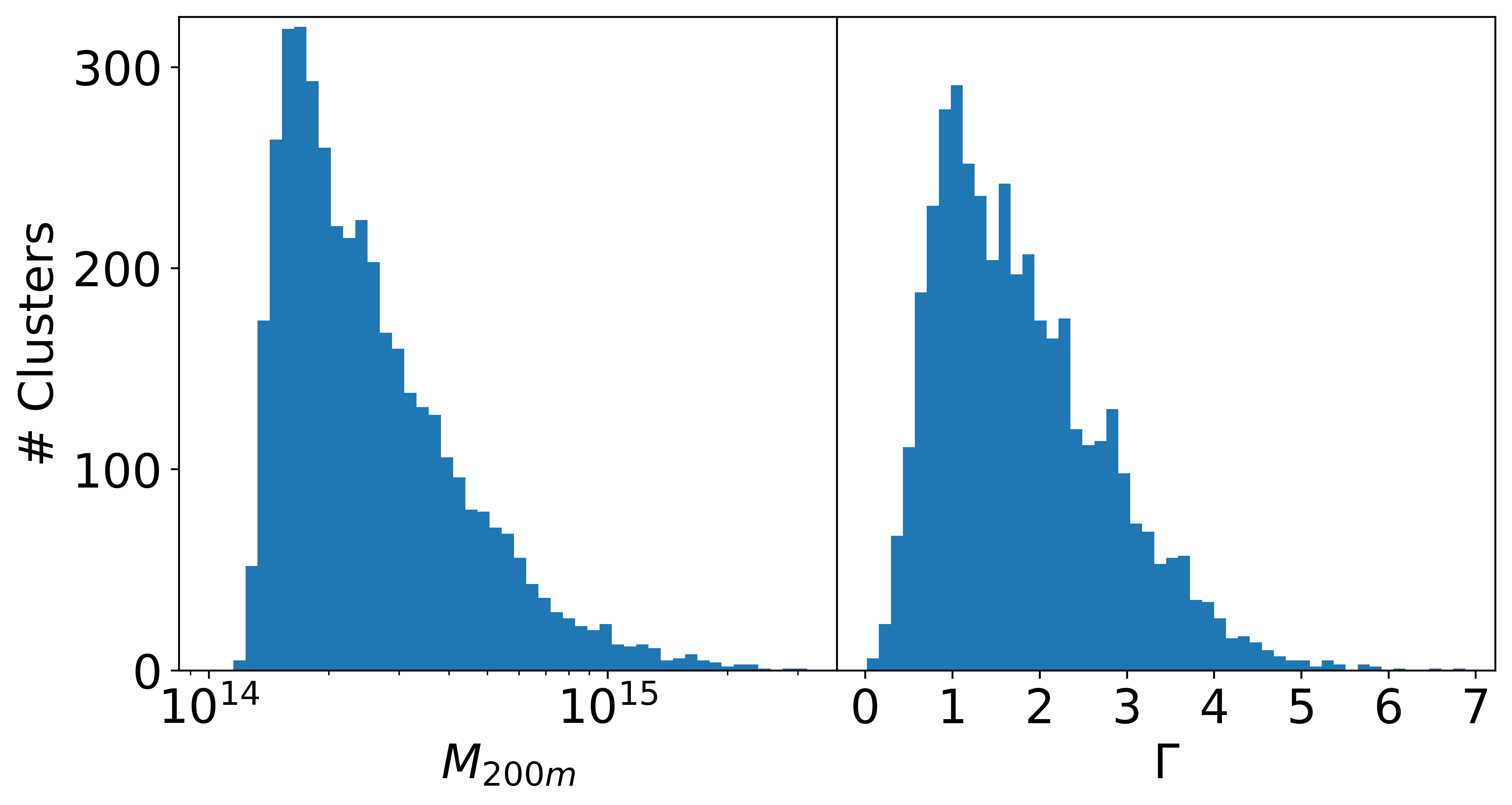}
    \caption{\raggedright Histogram of mass ($M_{200m}$) and MAR ($\Gamma$) values for the simulated galaxy clusters.  Both distributions are highly non-uniform.  To mitigate the biases that may be introduced by this non-uniformity, the MAR values are converted to z-score values for training. Further analyses of sources of bias is discussed in Section~\ref{biases}.}
    \label{fig:data_dist}
\end{figure}

The mock galaxy clusters were obtained from the MillenniumTNG (MTNG) hydrodynamic cosmological simulation\footnote{\url{https://www.mtng-project.org/}}. MTNG is a combination of the galaxy formation model of the IllustrisTNG simulation \citep{Nelson_2018, Pillepich_2018} and the very large Millennium dark matter only simulation \citep{Springel_2005}. It uses the $\Lambda$CDM cosmological parameter values from \citet{Planck_2015_Values}. The MTNG team has explained and analyzed their simulations in a series of introductory papers. In \citet{MTNG_1}, they perform an analysis of matter and halo statistics while introducing the technical aspects of the simulations. In \citet{MTNG_2}, the hydrodynamic simulations are described in more detail, and an analysis of the galaxy cluster population is performed. In \citet{MTNG_3}, the semi-analytic modeling code is updated and applied to produce lightcones for the dark-matter-only simulations. \citet{MTNG_4} examines the properties of the high redshift galaxy population. \citet{MTNG_6, MTNG_5} improve the halo occupation distribution models for the halo-galaxy connection. Galaxy clustering is analyzed in \citet{MTNG_7}. 
\citet{MTNG_8} explores cosmological parameter inference, while \citet{MTNG_9} analyzes the intrinsic alignment of galaxy shapes and large-scale structure. Finally, \citet{MTNG_10} studies the impact of baryons and massive neutrinos on weak lensing maps. We chose to use MTNG over other simulations because of the number of clusters available in the simulation; 4117 galaxy clusters with $M_{200c} \geq 10^{14} M_{\odot}$ at $z=0$ (see Figure \ref{fig:data_dist} for a visualization of the data distributions). 

For each cluster, we produce mock X-ray and tSZ observations. The X-ray observations were produced using the same procedure as in \citet{Ntampaka2019}, which we describe here in brief. 

The mock X-ray observations are produced using the \texttt{pyXSIM}\footnote{\url{http://hea-www.cfa.harvard.edu/~jzuhone/pyxsim/}} \citep{ZuHone2014} and \texttt{SOXS}\footnote{\url{http://hea-www.cfa.harvard.edu/soxs/}} software packages. Large photon samples are initially built in \texttt{pyXSIM} from the 3D density, temperature, and metallicity distributions of the MilleniumTNG data for each cluster using an APEC emission model \citep{Foster2012}, assuming a redshift of z = 0.05. Only particles with a temperature $T > 3 \times 10^5$~K and a gas density $\rho_g < 5 \times 10^{-25}$ g~cm$^{-3}$ that are not forming stars are used in the construction of the photon samples. These samples are then projected along each of the x-, y-, and z-axes of the simulation box, and foreground galactic absorption is applied to each sample assuming a \texttt{wabs} model \citep{wabs} with a value of $N_H = 1.8 \times 10^{20}$ cm$^2$.

These projected and absorbed events are then passed through the \texttt{SOXS} instrument simulator, which simulated 100~ks observations assuming the effective area and spectral response of {\it Chandra} as of its observing cycle 22, assuming 0.5" pixels and the {\it Chandra}-like PSF used in \texttt{SOXS}. The center of each observation is aimed at the cluster potential minimum. No backgrounds have been included in these mock observations. Each X-ray observation was separated into three bands: soft (0.5-1.2 keV), medium (1.2-2.0 keV), and hard (2.0-7.0 keV). 

For the tSZ mocks, we simply produce projected maps of the Compton-y parameter $y_{\rm tSZ}$ from our simulated clusters:
\begin{equation}
y_{\rm tSZ} = \int\frac{k_BT}{m_ec^2}\sigma_Tn_ed\ell,
\end{equation}
where $n_e$ is the electron number density and $\ell$ is the path length along the sight line. These tSZ observations are idealized, only including contributions from the simulated cluster, and do not include instrument response or background. 

The field of view of each observation was 39'. Given the redshift of $z=0.05$ from the observer, the corresponding angular diameter is$\sim1.2$ Mpc across. In our mock observations, this field of view is large enough to capture the full extent of most clusters in tSZ observations without having an excessive amount of empty pixels in the X-ray.

\begin{figure*}[tbh!]
    \centering
    \null \vspace{-5pt}
    \includegraphics[width=\textwidth]{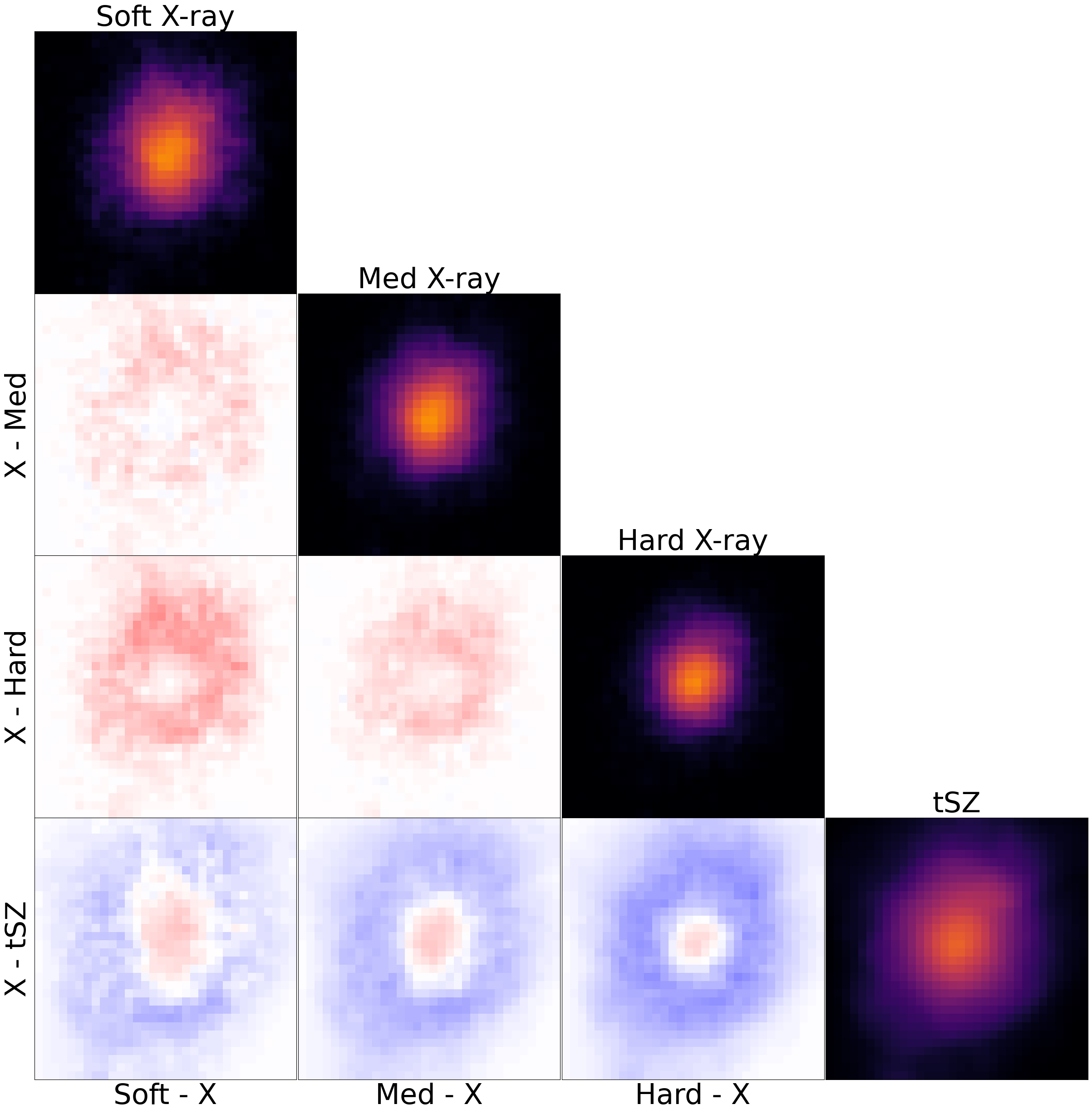}
    \caption{\raggedright A representative galaxy cluster mock observation across several energy bands. Each image at the top of a column is derived from the MTNG simulation and has been artificially lowered to 32x32 pixel resolution and then normalized following equation \ref{eqn:norm}. X-ray images show photon count in the corresponding energy bin, while tSZ image shows the Compton-y value. This example cluster has a true mass of $M_{200m}=2.76\times10^{14}M_{\odot}$ and a MAR of $\Gamma = 1.79$; from the distribution of masses and MARs in Figure \ref{fig:data_dist}, we see that this is a ``typical'' cluster. We also show the subtractions of each image to illustrate the spectral information available to the model. For more details about the simulation and mock observations, see Section \ref{data}. }
    \label{fig:sample_obs}
\end{figure*}

For computational reasons, each observation type was compressed into a 32$\times$32 pixel image. See Figure \ref{fig:sample_obs} for an example mock observation from our dataset. We briefly discuss the impact of varying the input data in Section~\ref{data_robustness}. 

The images were normalized so that the count per pixel value fell between 1 and 0 and did not range over many orders of magnitude. This was done to improve model performance. The images were normalized using:
\begin{equation}\label{eqn:norm}
    \tilde{X} = \tanh\left(\log_{10}\left(X/\bar{X} + 1\right)\right)
\end{equation}
where $X$ is the original 32x32 pixel image, $\bar{X}$ is the mean pixel value for that observation mode across the dataset, and $\tilde{X}$ is the normalized version of it. The multiplicative factor, $1/\bar{X}$, centers the peak of the distribution of pixel values.

For each galaxy cluster, we calculate the MAR according to the definition of \citet{Diemer_2020},
\begin{equation}\label{eqn:MAR_def}
    \Gamma_{\mathrm{dyn}} = \frac{\Delta \log M_{200m}}{\Delta \log a} \,.
\end{equation}
The time interval is the dynamical time, which we take to be the crossing time of the cluster at radius $R_{200m}$, written in units of the change in the logarithm of the scale factor $a$. This definition of MAR has been shown to most closely correlate with the dynamical state of the halo, and thus its splashback radius \citep{Shin_2023}. For further discussion of the relation between our choice of MAR and other definitions, see also \citet{Valles_Perez_2020}. Algorithmically, we calculate MARs using the \textsc{HydroTools} code \citep{diemer_17_sfh, diemer_18_hih2}, which follows the most massive progenitor history of each halo as tracked by the Subfind-HBT halo finder \citep{Springel_2021}. When training our model, we normalize our MARs by subtracting the mean MAR and dividing by their standard deviation, constructing a $z$-score. This normalization reduces the range of values and centers the values at zero, which improves the accuracy of the model. The distributions of MARs and mass values in our dataset are shown in Figure \ref{fig:data_dist}. 

\subsection{ML Model}\label{models}
Our model is a combination of a normalizing flows model \citep[see][for a review of normalizing flows]{Papamakarios_2019} and a convolutional neural network (CNN) model \citep[see][for a review of CNN and deep learning]{LeCun_2015}. We choose a CNN model to compress MAR-relevant information in the mock observations. We feed that information into the normalizing flows model, which allows us to perform posterior density estimation, giving us a measure of the uncertainty on the MAR estimate for each galaxy cluster. 

\begin{figure*}
    \centering
    %\null \vspace{-5pt}
    \includegraphics[width=\textwidth]{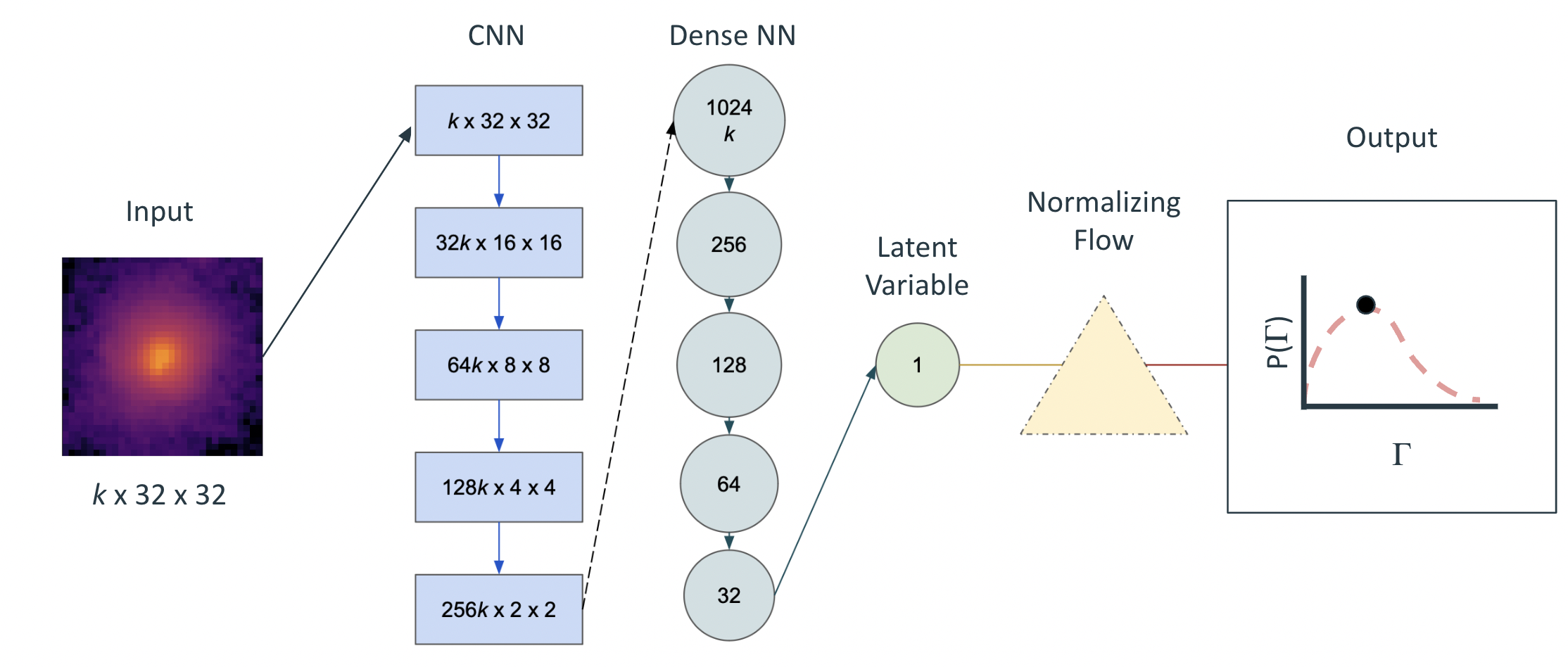}
    \caption{
    %\raggedright 
    A visualization of our ML model for an input with $k$ observation types. In the CNN portion of the model, images (blue boxes) are fed through a convolutional layer with a Leaky ReLU nonlinear activation function and a global pooling function (blue arrows). After four convolutional layers the final image is flattened out (black dashed arrow) into $1024 \times k$ weights. These are compressed through a series of fully connected layers (dark cyan arrows), each producing a smaller output (cyan circles). This culminates in a single latent space variable (green circle), which is then inputted into the normalizing flows model (yellow triangle) before outputting a sample MAR estimate from an approximation of the distribution of possible MARs given the input observation (far right).}
    \label{fig:model_diagram}
\end{figure*}

Our CNN model was written in the \texttt{Python} package \texttt{Pytorch}\footnote{\url{https://pytorch.org/}} and our normalizing flows model and training were implemented in the \texttt{Python} package \texttt{SBI}\footnote{\url{https://sbi-dev.github.io/sbi/}}. The CNN model serves as an embedding network, compressing the $4\times32\times32$ dimensional observation data into a single latent space variable. It does so utilizing convolution layers, which exploit locality and the translational invariance of features to efficiently compress information in images to higher levels of abstraction. 

Our CNN model consists of four convolution layers, the first starting with $32\times n_{\mathrm{obs}}$ filters, where $n_{\mathrm{obs}}$ is the number of observation types used (e.g., Soft X-ray only, or Soft X-ray and tSZ). Each subsequent convolution layer has twice as many filters as the previous layer. After each convolution layer, we apply a leaky ReLU layer \citep{LeakyReLU_2015} (unlike a standard ReLU, the $<0$ half has a non-zero slope) and then a two-dimensional max pooling layer, which reduces the dimensions of the image by a factor of two. At the end of the convolution layers, we are left with $256\times n_{\mathrm{obs}}$ images that are $2\times2$ pixels. We flatten this into a one-dimensional array, and four dense node layers follow. The first layer has $256\times n_{\mathrm{obs}}$ weights, and each subsequent dense layer has half as many. Between each layer, we once again apply a leaky ReLU non-linear activation function. After these dense layers, we apply one final dense layer, again with a leaky ReLU activation function, producing a single-scalar output, which we refer to as our latent space variable. The latent space variable is then inputted into the normalizing flows model. A diagram of our CNN model is shown in Figure \ref{fig:model_diagram}. It is possible that a different CNN model architecture could improve performance. While we did not aim to optimize our hyperparameters, we did explore a variety of possible model parameters, including increasing or decreasing the number of weights, changing the number of latent space variables that the CNN outputs, and removing max-pooling layers. We find that these changes did not improve model performance. The goal of our work is to demonstrate that such a model is possible and useful, not to provide a model ready for use on real data (see Section \ref{conclusion} for a further discussion concerning this point). 

The normalizing flows model trains a set of invertible transformations that map the data distribution (possible MARs of a cluster) to a base distribution (e.g., a normal distribution). By reversing these transformations, one can easily sample the base distribution, and thereby attain samples of the data distribution. A more detailed description of the exact procedure used to train our normalizing flows variant can be found in \citet{Greenberg_2019}. Our normalizing flows model, a SNPE-C type model \citep[][see also the \texttt{SBI} documentation]{Papamakarios_2018, Greenberg_2019}, uses the default model settings of the \texttt{SBI} package. We experimented with increasing the number of weights and transforms by a factor of two or more but found no significant improvement and increased computational demand. A diagram of our full model is shown in Figure \ref{fig:model_diagram}. 

\subsection{Fitting Function}\label{Diemer_2020}
To assess the effectiveness of our model, we compare its performance to the MAR fitting function of \citet{Diemer_2020}, which approximates MAR as defined in Equation \ref{eqn:MAR_def} as a function of mass and redshift,
\begin{equation}\label{eqn:Diemer_2020}
    \Gamma  = A\nu + B\nu^{3/2}
\end{equation}
\begin{equation}\label{eqn:Diemer_2020_A}
    A = 1.1721 + 0.3255z
\end{equation}
\begin{equation}\label{eqn:Diemer_2020_B}
    B  = -0.2565 + 0.0932z - 0.0571z^2 + 0.0042z^3 \,.
\end{equation}
Here, mass is expressed as peak height, $\nu$, which is a function of the $M_{200m}$, redshift, and cosmological parameters. $A$ and $B$ encode the redshift dependence through simple polynomial fits. We calculate the peak heights of each cluster using the \textsc{Colossus} code \citep{Diemer_2018} and use the transfer function from \citet{Eisenstein_1998} to approximate the power spectrum. The best-fit parameters of the fitting function are based on data from the Erebos suite of $N$-body simulations, which contains two different $\Lambda$CDM cosmologies \citep{Diemer_2020}. 

\begin{figure}%[b]
    \centering
    \null \vspace{-5pt}
    \includegraphics[width=3.2in]{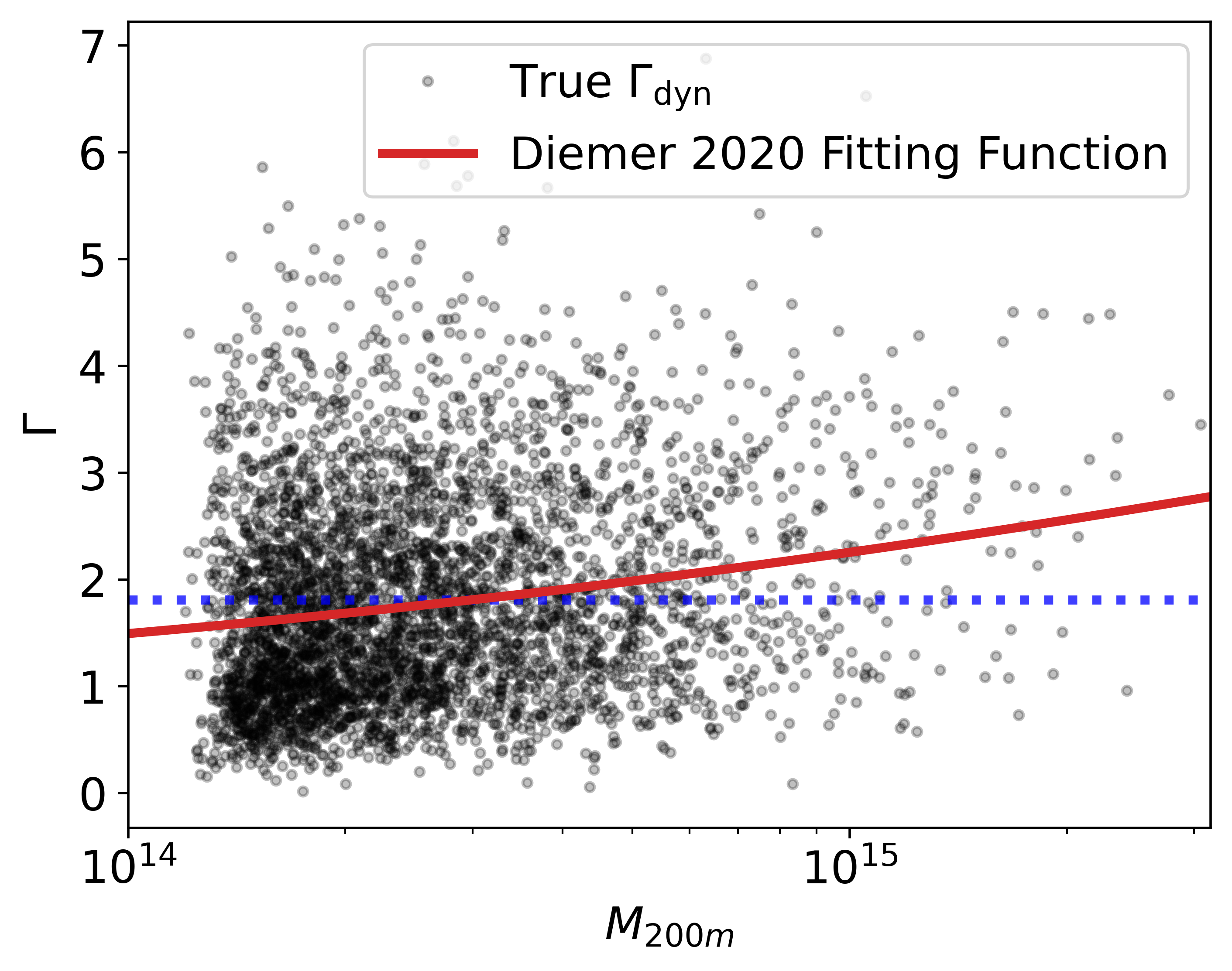}
    \caption{\raggedright A comparison of the fitting function to the true MAR for MTNG clusters in the dataset. The fitting function (red line), taken from \citep{Diemer_2020}, is a poor fit for the true MARs (black dots) of the MTNG galaxy clusters. For a fixed cosmology and redshift, the fitting function is solely a function of the galaxy cluster mass, which is weakly predictive of the MAR. Even so, the fitting function is the best existing non-ML method for estimating $\Gamma_{\mathrm{dyn}}$. The mean $\Gamma_{\mathrm{dyn}}$ (blue dotted line) is shown for comparison.}
    \label{fig:diemer_v_mass}
\end{figure}

Since our galaxy clusters are all from a single cosmology and a single redshift, the fitting function is only a function of mass (see Figure \ref{fig:diemer_v_mass}). The fitting function could be biased for a number of reasons including biases introduced by training on an $N$-body simulation and applying to a simulation with baryonic physics and biases from specific non-physical artifacts in the $N$-body code \citep[for a discussion of the robustness of MARs in $N$-body simulations see][]{Soltis_2024}. More importantly, the intrinsic scatter in MARs about any given galaxy cluster mass is high. The combination of these factors, but mostly the latter factor, means that the estimates of the fitting function are only weakly correlated to true MARs. A simple analysis reveals that the Pearson correlation coefficient \citep{Pearson_1895} between the estimates of the fitting function and the true MAR is $0.19$, while our ML model has a correlation coefficient of $0.77$ (for a more detailed discussion of the accuracy of the models, see Section \ref{results}).

\subsection{Training}\label{training}
The data is divided into tenths so that a 10-fold validation could be performed. Each tenth of data are assigned to a different test set, with the remaining data put in the validation and training set. Each fold contained a unique set of clusters, meaning that different lines of sight of the same cluster and rotations of a cluster observation were kept within the same data fold. The model training was handled by the \texttt{SBI} package\footnote{\url{https://sbi-dev.github.io/sbi/latest/reference/\#sbi.inference.snpe.snpe\_c.SNPE\_C.train}}. A batch size of 32 images was used. The default training function parameters were used. Training stopped when the validation loss failed to improve after 20 epochs. The typical training lasted $\sim30$ epochs. When analyzing our model's performance, we apply each model to its unique test set and then examine the joint set of estimates produced by all models.

In addition to training our model on all four possible observations of a given galaxy cluster (Soft X-ray, Medium X-ray, Hard X-ray, and tSZ), we also train to other variants on subsets of that data. The X-ray-only model is trained only on the X-ray observations of the clusters, and the tSZ-only model is trained only on the tSZ observations of the data. All other aspects of the training procedure and analysis are the same.

\section{Results}\label{results}
The principle advantage of using a normalizing flows model in our work is its capacity to estimate the posterior density, that is, the probability of the cluster having a given MAR given the observations of the cluster. We can then use this information to provide an estimate of the MAR and uncertainties for that estimate. For computational reasons, we approximate this procedure using 1000 samples from the trained machine-learning model. The median MAR estimate of the sample set is chosen as the fiducial estimate, and the 16th and 84th percentiles of the sample set are given as uncertainties. In the following subsections, we present the results of our model. In Section~\ref{accuracy}, we analyze the accuracy of the fiducial estimate and compare it to a fitting function. In Section~\ref{uncertainty}, we examine the accuracy of our uncertainty estimates. In Section~\ref{biases}, we test our fiducial estimate for biases. Finally, in Section~\ref{data_robustness}, we explore variants in the training data and model and what impact those choices have on model performance.

\subsection{Mass Accretion Rate Estimates}\label{accuracy}
To evaluate the overall accuracy of our ML model, we compare the accuracy of its estimates as compared to the accuracy of the fitting function. We do so by generating 1000 sample estimates for each cluster observation in the test set of each model, calculating the median of each sample set, and comparing that to the true MAR of the cluster, as determined using the procedure described in Section~\ref{data}. To make these results more interpretable, we convert the errors from units of MAR to percent error.

\begin{figure}%[b]
    \centering
    \null \vspace{-5pt}
    \includegraphics[width=3.2in]{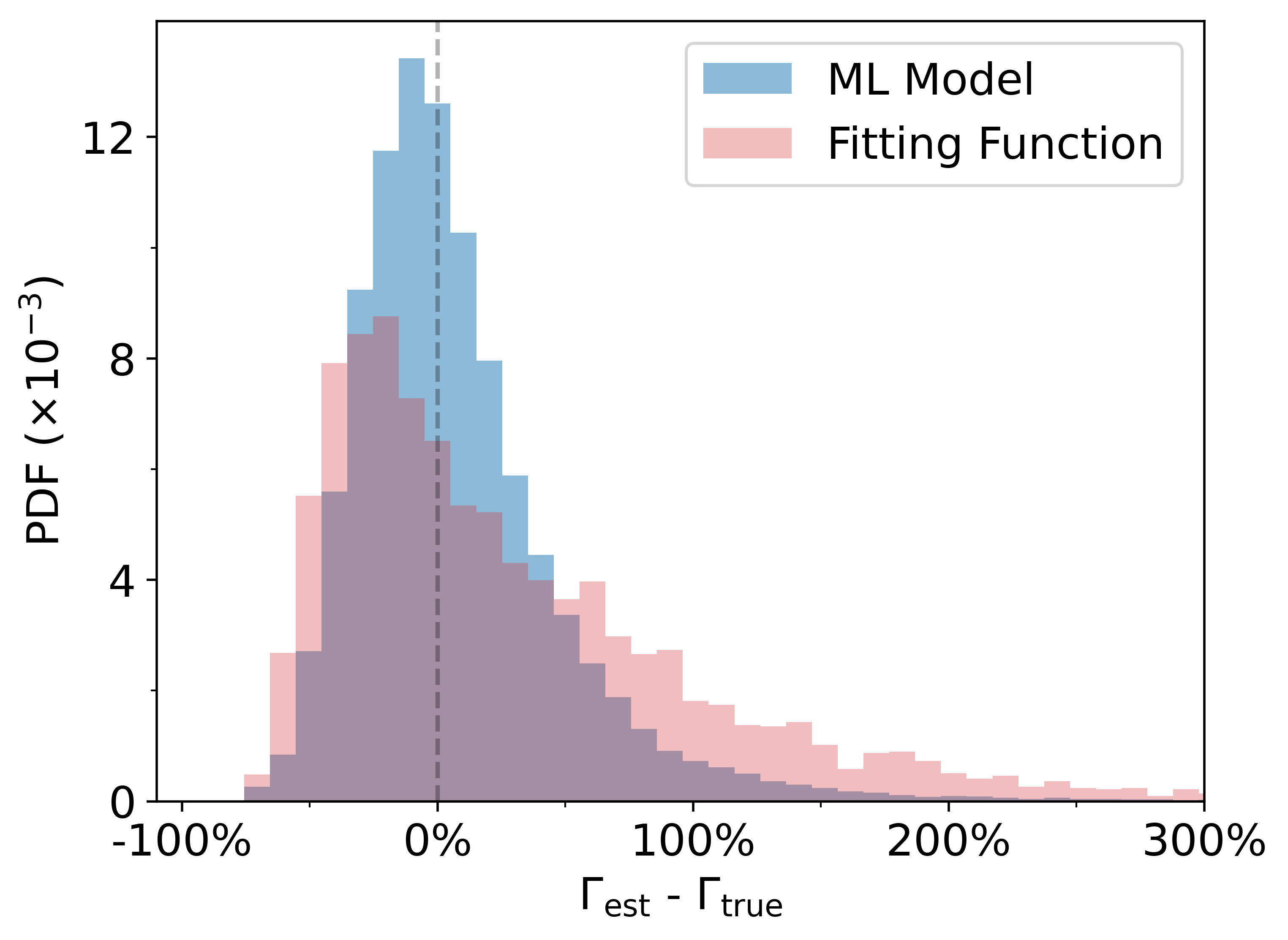}
    \caption{\raggedright Histogram of the percent error of the ML model MAR estimates (blue) compared to the fitting function (red). The ML model's error distribution has a smaller tail than the fitting function. For the ML model only 4\% of the cluster observations tested have MAR estimates with percent errors greater than 100\%. For the fitting function, this jumps to 16\% of the cluster observations tested. Similarly, more than half (56\%) of the observations tested have ML model estimates with less than 25\% error, but this reduces by a factor of 2 (33\%) for MAR estimates produced by the fitting function. A description of how the error is calculated can be found in Section~\ref{accuracy}.}
    \label{fig:percent_error}
\end{figure}

The resulting error distributions for the entire dataset are shown in Figure \ref{fig:percent_error}. Examining the absolute error, we find that our ML model outperforms the fitting function by nearly a factor of two. Moreover, we find that the use of a subset of the available observation modes does not substantially reduce the accuracy of the estimates. A tabular breakdown of our results is found in Table \ref{tab:median_est_errors}.

\begin{table}[]
    \centering
    \begin{tabular}{|c | c c c|} 
    \hline
    \textbf{MAR Estimator} & \textbf{50\%} & \textbf{68\%} & \textbf{95\%} \\
     & & & \\
    \hline
    \hline
    Fitting Function & 39\% & 56\% & 203\% \\ 
    \hline
    X-ray \& tSZ ML Model & 22\% & 33\% & 88\% \\
    \hline
    X-ray ML Model & 23\% & 35\% & 98\% \\
    \hline
    tSZ ML Model & 23\% & 34\% & 90\% \\
    \hline
    \end{tabular}
    \caption{The error of each estimator, for a given percent of the dataset. For example, for the MAR estimates produced by the fitting function, 50\% of the clusters in our dataset have a percent error of 39\% or less. The results show that the ML model out performs the fitting function by nearly a factor of two. Moreover, using either the X-ray observations alone, or the tSZ observations alone, does not result in a substantial loss of accuracy.}
    \label{tab:median_est_errors}
\end{table}

\subsection{Uncertainties}\label{uncertainty}
\begin{figure}%[b]
    \centering
    \null \vspace{-5pt}
    \includegraphics[width=3.2in]{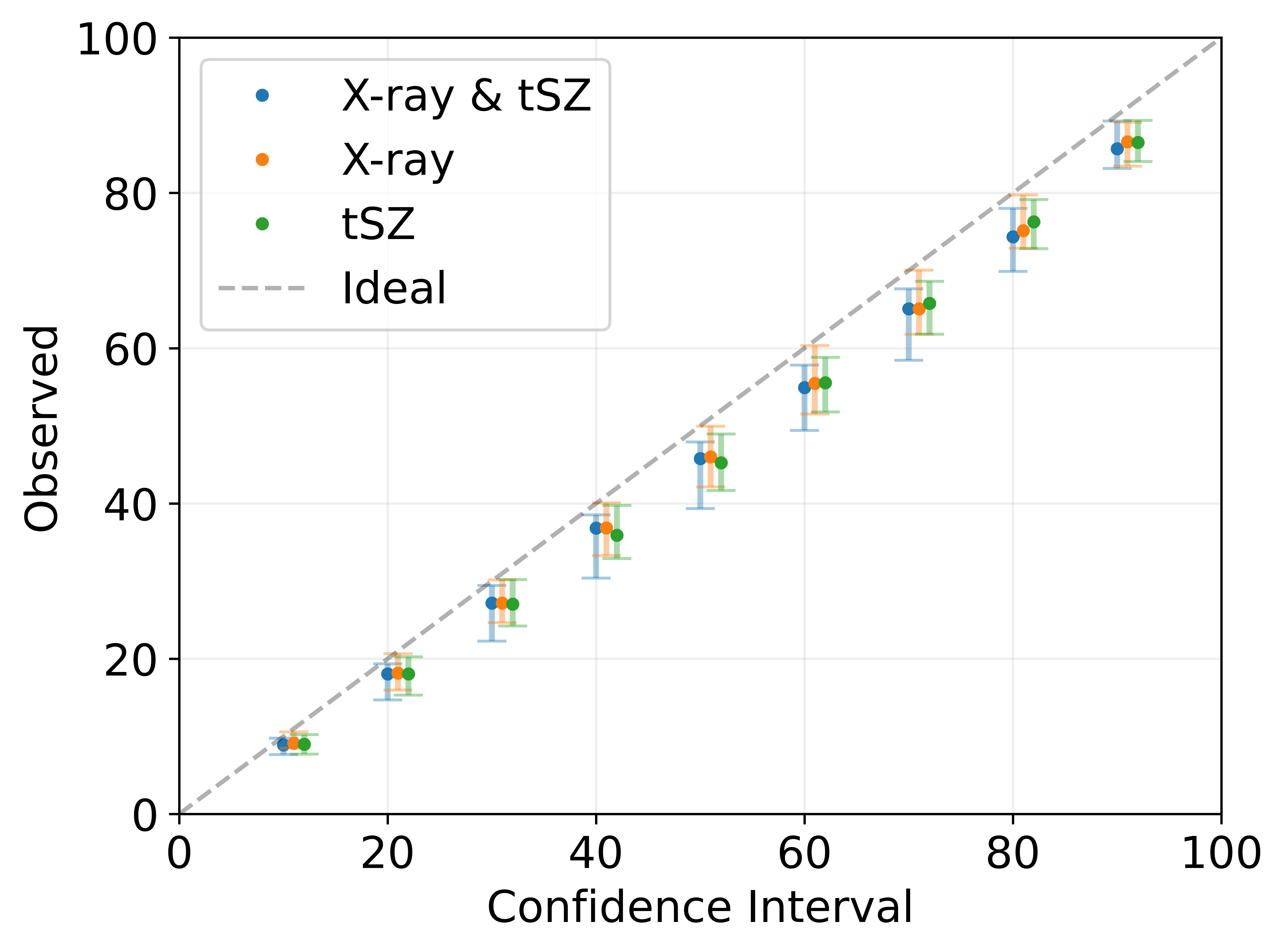}
    \caption{\raggedright Representation of the accuracy of the uncertainty estimates provided by the ML model. The model provides an accurate measure of the MAR estimate uncertainty. The median observed confidence intervals (points) are near to the case of perfectly calculated uncertainties (dashed line) for all models. The best-performing fold-unique model (top line of error bar) typically sits on the idealized line. We repeated this analysis ten times, using different data splits, and found the results were qualitatively similar. The error on the uncertainty estimates for all model variants are comparable. See Section \ref{uncertainty} for a detailed explanation of how the uncertainty estimate error is calculated. }
    \label{fig:uncertainties}
\end{figure}

How accurate are the uncertainty estimates produced by the model? We test this by examining how closely the model-generated confidence intervals match expectations. First, a single set of observations, we calculate the model-provided confidence interval using percentiles from the 1000 MAR estimate sample (see the introduction of Section \ref{results}). Second, we calculate whether the true MAR is contained within a given confidence interval (as calculated using the aforementioned percentiles). We repeat this procedure for the entire test set and for different potential confidence intervals. A visualization of the results is shown in Figure \ref{fig:uncertainties}. We find that the observed confidence interval, i.e., the frequency at which the true MAR is between the sample-derived percentiles, is a 1-5\% underestimate of the true uncertainty. We also find that using only X-ray or tSZ observations does not strongly impact the accuracy of the uncertainty estimates. We perform this evaluation over all fold-unique models (see Section \ref{training}), and find the variance in the observed confidence intervals ranges from 1-5\%. We repeated this analysis further by creating 10 more 10-fold divisions of the data, thereby training 100 new models in total. In evaluation of these new models, we found that the variance in the observed confidence intervals increased slightly, to a range $\sim$ 1-7\%. The worst underestimation in the model set underestimated the 50\% confidence interval by $11\%$. The most accurate observed confidence intervals were all within a percent of the ideal.

\subsection{Biases}\label{biases}
ML models are excellent with interpolation, but can struggle with extrapolation. As discussed in Section~\ref{data} and demonstrated in Figure \ref{fig:data_dist}, the mass and MAR distribution of the dataset is very non-uniform. From this information alone, we can suspect that the model is going to struggle to accurately estimate the MAR of clusters in the tails of the MAR distribution. If mass information is important for determining the MAR, it is also possible that the model will struggle with very low or very high mass clusters. 

\begin{figure}%[b]
    \centering
    \null \vspace{-5pt}
    \includegraphics[width=3.2in]{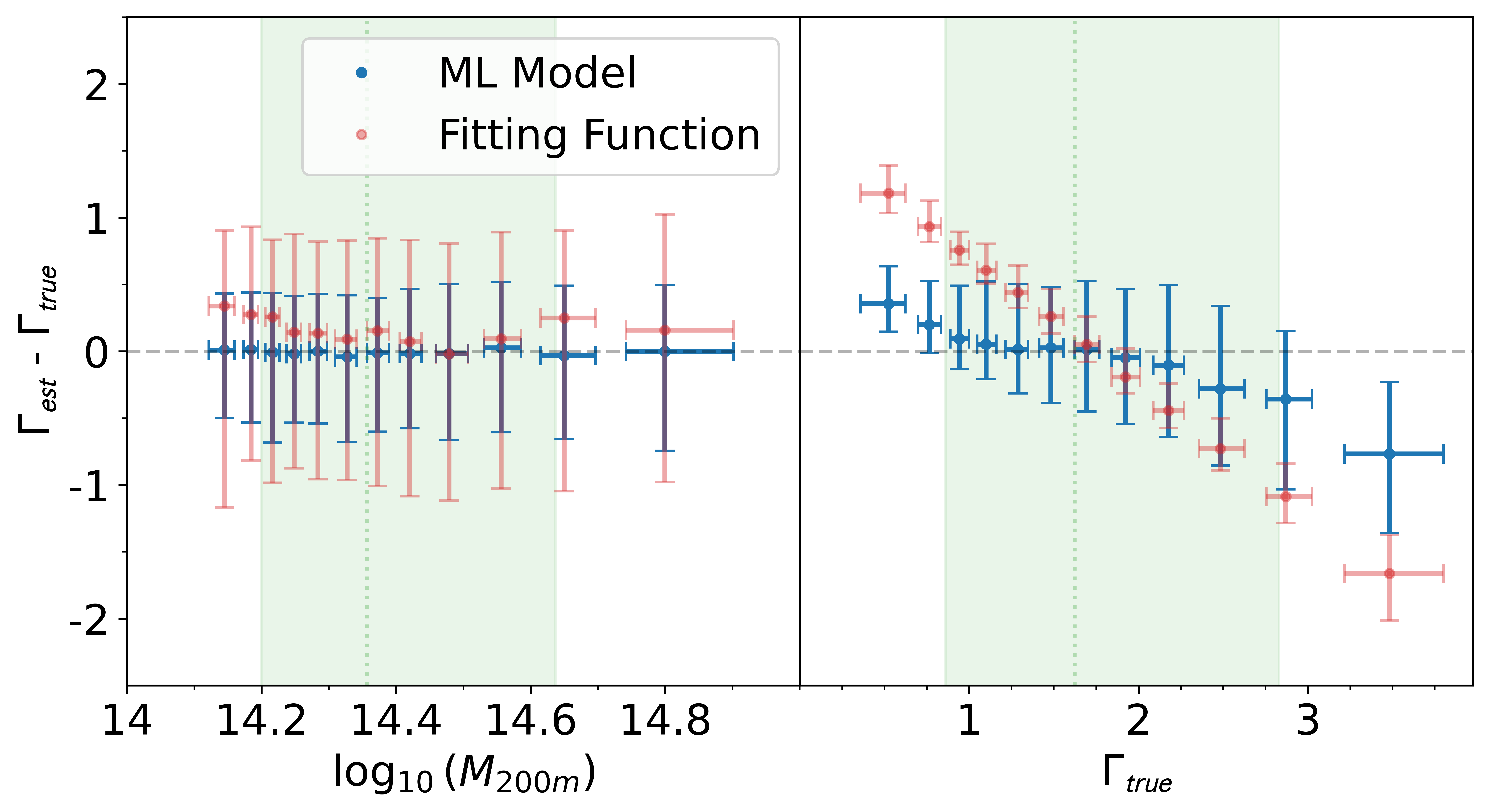}
    \caption{\raggedright Error bar plots comparing the ML model (blue) to the fitting function (red). All bins contain an equal number of clusters, with horizontal error bars showing the region within which 68\% of the clusters fall, and vertical error bars showing where 68\% of the errors fall (in units of $\Gamma$). The ML model displays no visible bias (points offset from the horizontal dashed line) as a function of galaxy cluster mass (left plot). The ML model is biased as a function of $\Gamma$ (right plot), but less so than the base line model. The bias increases as the cluster MAR strays from the median dataset MAR (vertical dotted line) and the region in which 68\% of the data falls (green shaded region). This bias is likely the result of the unbalanced training data, which contains few examples of very low or high MAR clusters.}
    \label{fig:mass_mar_bias}
\end{figure}

In order to test for these biases, we examined our ML model's median MAR estimate error as a function of mass and MAR. To investigate a potential mass dependency in our MAR estimate errors, we separate our data into 12 evenly populated bins, each monotonically increasing in mass. We then calculate the median error of each bin as well as the 16th and 84th percentile errors. We find that the median error in the bin and the scatter in the bin are consistent across mass steps. This result holds for the X-ray-only and tSZ-only models as well. For comparison, when this procedure is performed on the fitting function, we find that low mass ($\leq2.5\times 10^{14} M_{\odot}$) clusters and high mass ($\geq 5\times10^{14}$) clusters are more likely to have their MAR overestimated (see the right plot of Figure \ref{fig:mass_mar_bias}). The difference between the ML model and the performance of the fitting functions demonstrates that the ML model uses more than just the mass information to constrain the MAR.

To investigate a potential MAR-dependent bias, we perform a similar binning analysis using the true MAR values of the clusters. We find evidence of a bias in the model's MAR estimates, with low MAR clusters ($\Gamma<1$) having overestimated MARs and ($\Gamma>2$) having underestimated MARs. These biases reflect the distribution of true MAR in the dataset (see Figure \ref{fig:data_dist}). There is a long, narrow tail of very high MAR clusters in the distribution. The ML model is therefore underexposed to this region of parameter space and less effective at estimating these MARs. Nevertheless, even here, the model still outperforms the fitting function, which struggles to accurately estimate high MARs. It is worth noting that tracking progenitors along merger trees can sometimes lead to discontinuities in the mass history, occasionally leading to erroneously extreme MAR \citep[][see discussion in section 6]{halo_finder_comp}. While SUBFIND-HBT seeks to minimize this, it is possible some of the extreme MAR in our dataset are non-physical, which may hinder the model's accuracy. Finally, we again find that the behavior of the X-ray-only and tSZ-only models mirrors the X-ray and tSZ model, except that the X-ray and tSZ model tends to have smaller errors, and therefore smaller biases. The results of this and the mass dependence analysis are shown in Figure \ref{fig:mass_mar_bias}.

\subsection{Robustness to Data and Model Variants}\label{data_robustness}
While testing our model, we explored a variety of different input data formats (e.g., changes to the input observations). All of these changes had a minimal or negative impact on model performance. In addition to using \textit{Chandra}-like observations, we experimented with more idealized versions of X-ray observations. These idealized mock X-ray observations were surface brightness maps constructed directly from the simulations using the same emission model as the {\it Chandra}-like mocks. These were still split along the same energy bands but lacked any noise, background, or instrument response (unlike the \textit{Chandra}-like mocks). We found only minor improvements in the performance of the model. We also experimented with different image sizes (64x64 and 128x128), but found that no performance was lost when reducing to a smaller size, so we chose a smaller size to reduce computational demands. We believe that this consistency in accuracy across image sizes suggests that we do not have enough unique clusters in our dataset to fully exploit the relationship between small-scale features and MAR. We experimented with multiple different normalizations of the image data (see Equation \ref{eqn:norm}). In one variant, we removed the $tanh()$ function and in another we used a spline function to transform the pixel intensity distribution into a uniform distribution from 0 to 1. Neither of these normalizations improved model performance. 

When using a field-of-view that is fixed in physical size, it is possible for the ML model to learn something about the size of the cluster from its physical extent in the image. To test whether this was essential for estimating the MAR, we trained a model on fields-of-view that scaled depending on different definitions of the physical radius of galaxy clusters ($R_{500c}$, $R_{200m}$). We did not change the size of the images (all were 32x32). Models trained and tested on radius-dependent fields-of-view performed similarly to the fixed field-of-view model. When testing a model, trained on one field-of-view, on test data with a different field-of-view, the model accuracy breaks down, and estimates are no longer usable. When a model is trained on multiple fields-of-view simultaneously and then tested on those fields-of-view, the model achieves a similar accuracy on each field-of-view as the models analyzed in Section~\ref{accuracy}, but training is less stable and uncertainties provided by the model are more severely underestimated. Moreover, training on multiple fields-of-view simultaneously results in a larger training set, thereby increasing the computational cost. Consistent accuracy across models trained with different field-of-view images suggests that the necessary information for MAR determination exists at many length scales and radial distances from the core, or that the important information is primarily global, and that the model is not simply using the physical extent of the cluster to constrain MAR.

We also experimented with model variants. In particular, we were interested in the impacts of changing the size of the latent space that connects the CNN and normalizing flows components of the model. We found that increasing the latent space to 5 or 10 variables did not result in a significant change in the performance of the model. Factor of two changes to the size of the model (i.e., the number of weights and CNN filters) did not result in any substantial changes in model performance. Since increasing the model size increased the computational demands for training the model without improving performance, we opted for the aforementioned model architecture.

\section{Image Decomposition \& Interpretation}\label{interpretation}

\begin{figure*}%[b]
    \centering
    \null \vspace{-5pt}
    \includegraphics[width=\textwidth]{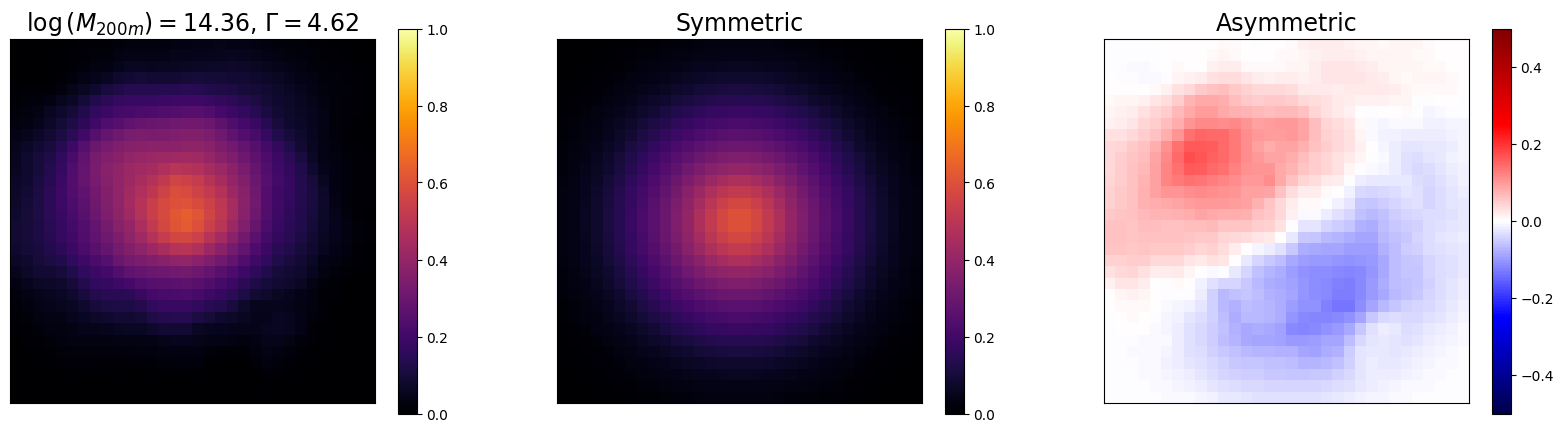}
    \caption{\raggedright A mock tSZ observation of a high MAR galaxy cluster (left) decomposed into symmetric (center) and asymmetric (right) components. The image decomposition technique, discussed in Section~\ref{interpretation}, splits images into radially symmetric and asymmetric components. Using this technique  we analyze the relative importance of the radial density profile and the asymmetry of the cluster for MAR estimates.}
    \label{fig:image_decomp_example}
\end{figure*}

We expect the MAR to influence the morphology of galaxy clusters in different ways. From \citet{Diemer_2014} we know that MAR influences the radial density profile of the cluster. However, we also expect that the MAR influences the ellipticity and the fraction of mass contained in the substructure \citep[e.g.,][]{Wong_2012, Jiang_2016}. While our images are very low resolution, all of these pieces of information are present to some degree. To better understand what features in the image influence model estimates, we decompose each cluster into two components, a radially symmetric component and an asymmetric component. The radially symmetric component isolates the importance of the radial density profile of the cluster in MAR determination. The asymmetry component contains a combination of substructure and ellipticity information. By studying the dependence of model estimates on changes in the asymmetric and symmetric components, we hope to better understand the relationship between MAR and the features the components represent.

To decompose the images, we first calculate a radial profile of each image using a one-dimensional Gaussian density kernel with a standard deviation of 0.6. We chose a smoothing factor that was smaller than a single pixel to avoid excessively distorting the true profile. We then correct for pixelation-related artifacts in the profile by dividing by the radial profile of a $32\times32$ matrix of ones\footnote{We know that the profile of an image of ones must be one, therefore any deviation from one is a pixelation error. Dividing by this uncorrected profile removes these pixelation-related errors.}. We used the corrected radial profile to construct a two-dimensional image, calculating the value of each pixel using its distance from the center of the image. The asymmetric image component is the original image minus the symmetric image. We do this calculation for all observation modes and do so on only normalized images. An example of this procedure is shown in Figure \ref{fig:image_decomp_example}.

\begin{figure*}%[b]
    \centering
    \null \vspace{-5pt}
    \includegraphics[width=\textwidth]{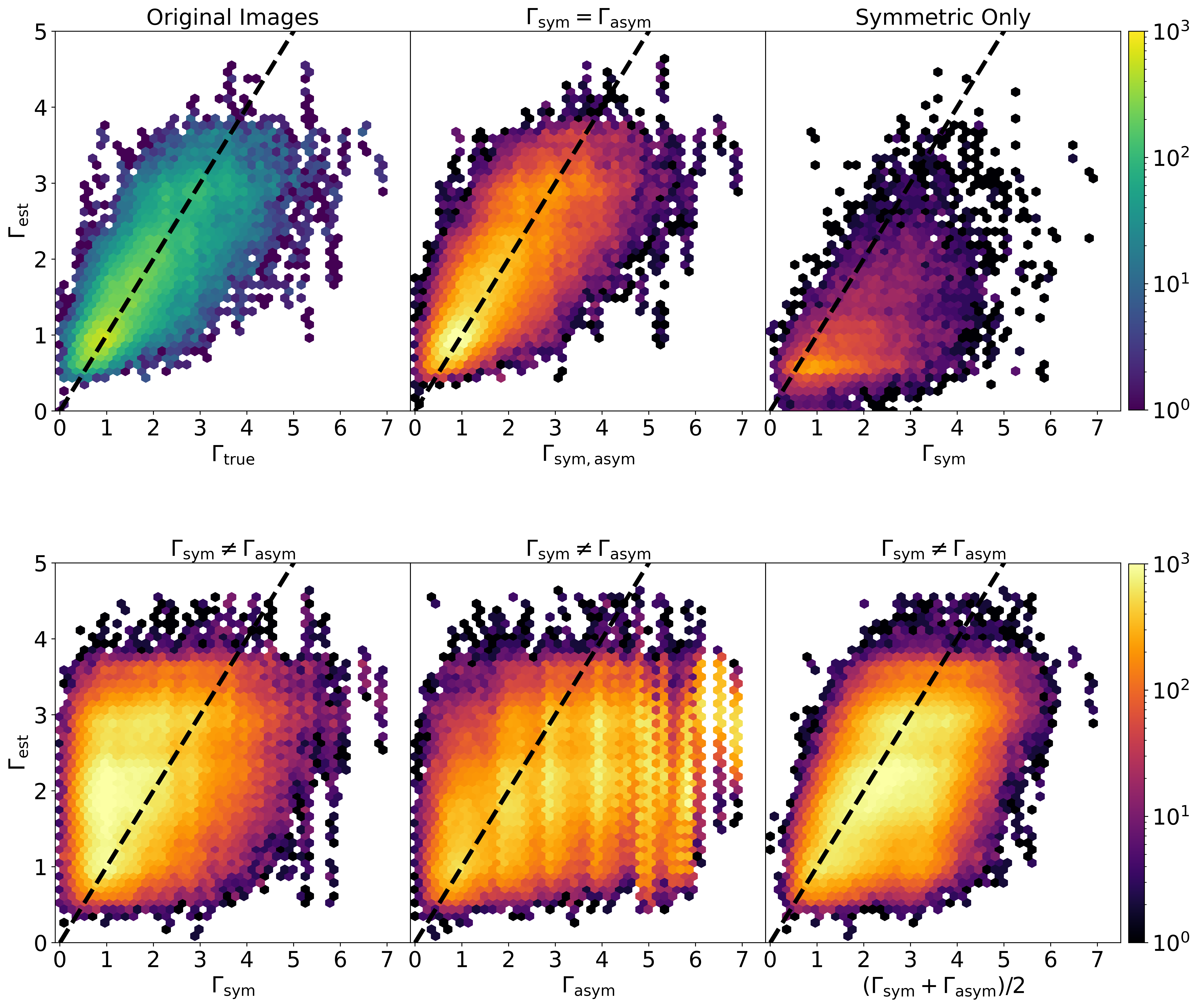}
    \caption{\raggedright 2d histograms of the estimated MARs vs true MARs for different combinations of radially symmetric and asymmetric components. The top left plot, shown with its own color mapping, is the models performance on the unmodified mock observations. The model accuracy is not greatly diminished by combining asymmetric and symmetric components from different lines-of-sight of the same cluster (top center as compared to top left, $\Gamma_{sym} = \Gamma_{asym}$ denotes that both components were calculated from observations of the same cluster). The top right plot shows the model performance when shown only the symmetric components, which results in a persistent underestimate of MAR, suggesting the asymmetric image component is essential to model estimates. When different clusters' components are combined, the MAR estimated is not strongly correlated with the symmetric component (bottom left) or the asymmetric component (bottom center). The correlation between estimates and the average of the MARs of the symmetric and asymmetric components again suggests that both components are essential.}
    \label{fig:symmetric_vs_asymmetric}
\end{figure*}

In our first test, we combined only symmetric and asymmetric components of the same cluster. The radial profile and the asymmetric component of a cluster change depending on the line-of-sight. By mixing and matching the lines-of-sight of the same cluster, we can test how sensitive the model is to these changes. The results of our test are shown in the top center plot of Figure \ref{fig:symmetric_vs_asymmetric}. Using our X-ray and tSZ combined model, we find that the MAR estimates are nearly as accurate as when the original images were used. Using the same accuracy metric as in Section~\ref{accuracy}, median estimate percent error percentiles, we find the bounds on the percent errors for 50\%, 68\%, and 95\% of the data are 22\%, 34\%, and 94\%. These results are nearly identical to the model applied to unmodified observations, which strongly suggests that the model is not using information that requires both the symmetric and asymmetric components to have the same line-of-sight. The differences in the radial profiles of a cluster across different lines-of-sight, which are smaller than differences in radial profiles between clusters, are likely to be too small to impact model performance.

In our second test, we use only symmetric inputs. This test allows us to isolate if the model is using asymmetric components at all, and if so, what impact their presence has on model estimates. Naively, we would assume that a radially symmetric cluster is relaxed and therefore has a lower MAR than one with apparent asymmetries. This is indeed what we find, which can be seen in the top right plot of Figure \ref{fig:symmetric_vs_asymmetric}. The model systematically underestimates the MAR of clusters when given only their symmetric components. This suggests that the model uses the asymmetric components, which contain both ellipticity and substructure information. This was further explored by training and testing versions of the model on the symmetric and asymmetric components separately. The results suggest that both the symmetric and asymmetric components contain sufficient information to estimate MAR, but that the accuracy achieved by the model presented requires information from both components.

In our third test, we combine the symmetric and asymmetric components of different clusters of similar mass. We then input these ``Frankenstein" clusters into the ML model and obtain MAR estimates for each. We can compare this MAR estimate to the MAR of the cluster used to produce the symmetric component and to the MAR of the cluster used to produce the asymmetric component. The goal of this test is two-fold. Firstly, to determine whether either component is dominant in influencing the MAR estimate. Secondly, to determine whether there is information contained in the union of the symmetric and asymmetric components.

For this test, we separated the clusters into logarithmically spaced mass bins, going from $\log M_{200m} = 14$ to $\log M_{200m} = 16$ in steps of $\Delta\log M_{200m} = 0.25$. We only combined components from clusters that fell within the same mass bin. We did this so that the total size of the cluster in the image would roughly match. We also binned the clusters by their MAR, going from $\Gamma = 0$ to $\Gamma 
 = 7$ in steps of $\Delta \Gamma = 1$. We added the symmetric component of each cluster to the asymmetric component of a different cluster within the same mass bin. We repeated this seven times, so that each cluster was combined with another cluster from the same mass bin and from each of the seven MAR bins. We then compared the predictions of the ML model with the true MAR of the cluster seen in the symmetric component, the true MAR seen in the asymmetric component, and the average of the true MARs of the clusters in both components. The results of the test for mismatched components are shown in the bottom row of Figure \ref{fig:symmetric_vs_asymmetric}. MAR estimates produced by the model are not strongly correlated with either the symmetric or asymmetric component. Comparing the averages of the symmetric and asymmetric MAR improves the correlation somewhat, but not significantly.

 Our third test allows us to see the impact of the MAR of the asymmetric component, controlling for the mass and the MAR of the symmetric component. We find that there is a small mild positive correlation between $\Gamma_{\rm est} - \Gamma_{\rm sym}$ and $\Gamma_{\rm asym} - \Gamma_{\rm sym}$, where $\Gamma_{\rm est}$ is the MAR of the model, $\Gamma_{\rm sym}$ is the MAR of the symmetric component cluster, and $\Gamma_{\rm asym}$ is the MAR of the asymmetric component cluster. In other words, the model estimates a higher MAR relative to the MAR of the symmetric component when a higher MAR asymmetric component is added (and vice versa). This trend is strongest for low-mass clusters and low-MAR symmetric components, but is mild even in the strongest cases. The weakness of the trend suggests that there is important information contained in the union of the symmetric and asymmetric components that is lost when the cluster components are mixed. This is further confirmed by the high scatter relationship between $\Gamma_{\rm est}$ and $\Gamma_{\rm sym}$ when $\Gamma_{\rm asym} \simeq \Gamma_{\rm sym}$.

\section{Conclusion}\label{conclusion}
We have developed a MAR estimation technique that uses X-ray and tSZ observations, in combination or individually, to constrain the MAR of mock-observed clusters. This ML model is nearly a factor of two more precise than mass-based estimates, the next best available technique. Our mock observations rely on simplifying assumptions, and applying this method to real cluster observations will introduce further challenges. Our model was trained using a single hydrodynamic cosmological simulation, with a single set of assumptions about astrophysics (e.g., AGN feedback) and cosmology. A future model applied to real data will need to be trained on simulated observations with varying astrophysical and cosmological assumptions to reduce the dependence of MAR estimates on these assumptions. Furthermore, our simulated clusters are all observed at the same point in time ($z=0$) and from the same distance ($z=0.05$). A more realistic data set will need to include galaxy clusters observed at different redshifts, to capture differences in structure formation and in foreground (including foreground clusters and filaments). Furthermore, a future model will need to use more realistic observations, perhaps mimicking potential future X-ray experiments (e.g., the Line Emission Mapper\footnote{\url{https://www.lem-observatory.org/}} and Athena\footnote{\url{https://www.the-athena-x-ray-observatory.eu}}). Future work might also explore the benefits of incorporating additional observation modes (e.g., optical, spectral, weak lensing, and radio observations), as well as modifying the ML model architecture.

Interpreting neural networks is difficult, but we are able to learn useful information in doing so. We verify that our model does not simply use the mass information of the cluster to constrain MAR, as that information is insufficient to achieve our model accuracy. Similarly, experimenting with different fields-of-view confirms that MAR information exists across different length scales and radii. Decomposing images into symmetric and asymmetric components indicates that both the cluster profile and cluster asymmetry are essential to obtain the level of accuracy of our model. The radial profile alone, if used to produce a radially symmetric image, will lead to an underestimate of the MAR. Moreover, there is important information contained in the union of the symmetric and asymmetric components that is essential to our model's performance.

Galaxy cluster MARs must be better understood if we are to answer crucial questions about galaxy cluster physics, improve galaxy cluster mass count constraints of cosmology, or create a new probe of cosmology using MARs. Our ML model, which combines a CNN embedding network with a normalizing flows model, is capable of estimating the MAR of clusters nearly a factor-of-two better than using mass to estimate MAR, while also simultaneously providing an estimate of the uncertainty on those constraints. Further work on directly constraining galaxy cluster MAR offers a promising route towards understanding the galaxy cluster mass bias, improving cosmological parameter constraints derived from galaxy clusters, and enhancing our understanding of the astrophysics that govern the internal dynamics of galaxy clusters.

\acknowledgements We thank Matthew Ho, Arya Farahi, Philip Mansfield, Tibor Rothschild, Ismael Mendoza, Frank van den Bosch, Naomi Gluck, and Priyamvada Natarajan for conversations, questions, and suggestions that led to improvements in this work. 
The material presented is based on work supported by NASA under award No. 80NSSC22K0821.  This work was carried out at the Advanced Research Computing at Hopkins (ARCH) core facility  (rockfish.jhu.edu), which is supported by the National Science Foundation (NSF) grant number OAC 1920103. Ana Maria Delgado acknowledges support from NSF grant number 2307070. CH-A acknowledges support from the Excellence Cluster ORIGINS funded by
the Deutsche Forschungsgemeinschaft (DFG, German Research Foundation)
under Germany's Excellence Strategy -- EXC-2094 -- 390783311. SB acknowledges funding from a UK Research \& Innovation (UKRI) Future Leaders Fellowship [grant number MR/V023381/1]. 

\bibliography{myref.bib}

\begin{thebibliography}{}
\expandafter\ifx\csname natexlab\endcsname\relax\def\natexlab#1{#1}\fi

\bibitem[{{Amoura} {et~al.}(2024){Amoura}, {Drakos}, {Berrouet}, \&
  {Taylor}}]{Amoura_2024}
{Amoura}, Y., {Drakos}, N.~E., {Berrouet}, A., \& {Taylor}, J.~E. 2024, \mnras,
  527, 3459

\bibitem[{{Andrade} {et~al.}(2022){Andrade}, {Fuson}, {Gad-Nasr}, {Kong},
  {Minor}, {Roberts}, \& {Kaplinghat}}]{Andrade_2022}
{Andrade}, K.~E., {Fuson}, J., {Gad-Nasr}, S., {et~al.} 2022, \mnras, 510, 54

\bibitem[{{Arendt} {et~al.}(2024){Arendt}, {Perrott}, {Contreras-Santos}, {de
  Andres}, {Cui}, \& {Rennehan}}]{Arendt_2024}
{Arendt}, A.~R., {Perrott}, Y.~C., {Contreras-Santos}, A., {et~al.} 2024,
  \mnras, 530, 20

\bibitem[{{Barrera} {et~al.}(2023){Barrera}, {Springel}, {White},
  {Hern{\'a}ndez-Aguayo}, {Hernquist}, {Frenk}, {Pakmor}, {Ferlito},
  {Hadzhiyska}, {Delgado}, {Kannan}, \& {Bose}}]{MTNG_3}
{Barrera}, M., {Springel}, V., {White}, S. D.~M., {et~al.} 2023, \mnras, 525,
  6312

\bibitem[{{Bleem} {et~al.}(2024){Bleem}, {Klein}, {Abbot}, {Ade}, {Aguena},
  {Alves}, {Anderson}, {Andrade-Oliveira}, {Ansarinejad}, {Archipley}, {Ashby},
  {Austermann}, {Bacon}, {Beall}, {Bender}, {Benson}, {Bianchini}, {Bocquet},
  {Brooks}, {Burke}, {Calzadilla}, {Carlstrom}, {Carnero Rosell}, {Carretero},
  {Chang}, {Chaubal}, {Chiang}, {Chou}, {Citron}, {Corbett Moran}, {Costanzi},
  {Constanzi}, {Crawford}, {Crites}, {da Costa}, {de Haan}, {De Vicente},
  {Desai}, {Dobbs}, {Doel}, {Everett}, {Ferrero}, {Flaugher}, {Floyd},
  {Friedel}, {Frieman}, {Gallicchio}, {Garc'ia-Bellido}, {Gatti}, {George},
  {Giannini}, {Grandis}, {Gruen}, {Gruendl}, {Gupta}, {Gutierrez}, {Halverson},
  {Hinton}, {Hinton}, {Holder}, {Hollowood}, {Holzapfel}, {Honscheid},
  {Hrubes}, {Huang}, {Hubmayr}, {Irwin}, {Mena-Fern{\'a}ndez}, {James},
  {K{\'e}ruzor{\'e}}, {Knox}, {Kuehn}, {Lahav}, {Lee}, {Lee}, {Li}, {Lowitz},
  {Marshal}, {McDonald}, {McMahon}, {Menanteau}, {Meyer}, {Miquel}, {Mohr},
  {Montgomery}, {Myles}, {Natoli}, {Nibarger}, {Noble}, {Novosad}, {Ogando},
  {Padin}, {Patil}, {Pereira}, {Pieres}, {Plazas Malag'on}, {Pryke},
  {Reichardt}, {Rodr'iguez-Monroy}, {Romer}, {Ruhl}, {Saliwanchik}, {Salvati},
  {Sanchez}, {Saro}, {Schaffer}, {Schrabback}, {Sevilla-Noarbe}, {Sievers},
  {Smecher}, {Smith}, {Somboonpanyakul}, {Stalder}, {Stark}, {Suchyta},
  {Swanson}, {Tarle}, {To}, {Tucker}, {Veach}, {Vieira}, {Vincenzi}, {Wang},
  {Weller}, {Whitehorn}, {Wiseman}, {Wu}, {Yefremenko}, {Zebrowski}, \&
  {Zhang}}]{SPT_catalog}
{Bleem}, L.~E., {Klein}, M., {Abbot}, T.~M.~C., {et~al.} 2024, The Open Journal
  of Astrophysics, 7, 13

\bibitem[{{Bocquet} {et~al.}(2024){Bocquet}, {Grandis}, {Bleem}, {Klein},
  {Mohr}, {Schrabback}, {Abbott}, {Ade}, {Aguena}, {Alarcon}, {Allam}, {Allen},
  {Alves}, {Amon}, {Anderson}, {Annis}, {Ansarinejad}, {Austermann}, {Avila},
  {Bacon}, {Bayliss}, {Beall}, {Bechtol}, {Becker}, {Bender}, {Benson},
  {Bernstein}, {Bhargava}, {Bianchini}, {Brodwin}, {Brooks}, {Bryant},
  {Campos}, {Canning}, {Carlstrom}, {Carnero Rosell}, {Carrasco Kind},
  {Carretero}, {Castander}, {Cawthon}, {Chang}, {Chang}, {Chaubal}, {Chen},
  {Chiang}, {Choi}, {Chou}, {Citron}, {Corbett Moran}, {Cordero}, {Costanzi},
  {Crawford}, {Crites}, {da Costa}, {Pereira}, {Davis}, {Davis}, {DeRose},
  {Desai}, {de Haan}, {Diehl}, {Dobbs}, {Dodelson}, {Doux}, {Drlica-Wagner},
  {Eckert}, {Elvin-Poole}, {Everett}, {Everett}, {Ferrero}, {Fert{\'e}},
  {Flores}, {Frieman}, {Gallicchio}, {Garc{\'\i}a-Bellido}, {Gatti}, {George},
  {Giannini}, {Gladders}, {Gruen}, {Gruendl}, {Gupta}, {Gutierrez},
  {Halverson}, {Harrison}, {Hartley}, {Herner}, {Hinton}, {Holder},
  {Hollowood}, {Holzapfel}, {Honscheid}, {Hrubes}, {Huang}, {Hubmayr}, {Huff},
  {Huterer}, {Irwin}, {James}, {Jarvis}, {Khullar}, {Kim}, {Knox}, {Kraft},
  {Krause}, {Kuehn}, {Kuropatkin}, {K{\'e}ruzor{\'e}}, {Lahav}, {Lee}, {Leget},
  {Li}, {Lin}, {Lowitz}, {MacCrann}, {Mahler}, {Mantz}, {Marshall},
  {McCullough}, {McDonald}, {McMahon}, {Mena-Fern{\'a}ndez}, {Menanteau},
  {Meyer}, {Miquel}, {Montgomery}, {Myles}, {Natoli}, {Navarro-Alsina},
  {Nibarger}, {Noble}, {Novosad}, {Ogando}, {Omori}, {Padin}, {Pandey},
  {Paschos}, {Patil}, {Pieres}, {Plazas Malag{\'o}n}, {Porredon}, {Prat},
  {Pryke}, {Raveri}, {Reichardt}, {Roberson}, {Rollins}, {Romero}, {Roodman},
  {Ruhl}, {Rykoff}, {Saliwanchik}, {Salvati}, {S{\'a}nchez}, {Sanchez},
  {Sanchez Cid}, {Saro}, {Schaffer}, {Secco}, {Sevilla-Noarbe}, {Sharon},
  {Sheldon}, {Shin}, {Sievers}, {Smecher}, {Smith}, {Somboonpanyakul},
  {Sommer}, {Stalder}, {Stark}, {Stephen}, {Strazzullo}, {Suchyta}, {Tarle},
  {To}, {Troxel}, {Tucker}, {Tutusaus}, {Varga}, {Veach}, {Vieira},
  {Vikhlinin}, {von der Linden}, {Wang}, {Weaverdyck}, {Weller}, {Whitehorn},
  {Wu}, {Yanny}, {Yefremenko}, {Yin}, {Young}, {Zebrowski}, {Zhang}, {Zohren},
  \& {Zuntz}}]{SPT_Abundances}
{Bocquet}, S., {Grandis}, S., {Bleem}, L.~E., {et~al.} 2024, arXiv e-prints,
  arXiv:2401.02075

\bibitem[{{Bose} {et~al.}(2023){Bose}, {Hadzhiyska}, {Barrera}, {Delgado},
  {Ferlito}, {Frenk}, {Hern{\'a}ndez-Aguayo}, {Hernquist}, {Kannan}, {Pakmor},
  {Springel}, \& {White}}]{MTNG_7}
{Bose}, S., {Hadzhiyska}, B., {Barrera}, M., {et~al.} 2023, \mnras, 524, 2579

\bibitem[{{Bulbul} {et~al.}(2024){Bulbul}, {Liu}, {Kluge}, {Zhang}, {Sanders},
  {Bahar}, {Ghirardini}, {Artis}, {Seppi}, {Garrel}, {Ramos-Ceja}, {Comparat},
  {Balzer}, {B{\"o}ckmann}, {Br{\"u}ggen}, {Clerc}, {Dennerl}, {Dolag},
  {Freyberg}, {Grandis}, {Gruen}, {Kleinebreil}, {Krippendorf}, {Lamer},
  {Merloni}, {Migkas}, {Nandra}, {Pacaud}, {Predehl}, {Reiprich}, {Schrabback},
  {Veronica}, {Weller}, \& {Zelmer}}]{Bulbul_2024}
{Bulbul}, E., {Liu}, A., {Kluge}, M., {et~al.} 2024, \aap, 685, A106

\bibitem[{{Chen} {et~al.}(2019){Chen}, {Avestruz}, {Kravtsov}, {Lau}, \&
  {Nagai}}]{Chen_2019}
{Chen}, H., {Avestruz}, C., {Kravtsov}, A.~V., {Lau}, E.~T., \& {Nagai}, D.
  2019, \mnras, 490, 2380

\bibitem[{{Contreras} {et~al.}(2023){Contreras}, {Angulo}, {Springel}, {White},
  {Hadzhiyska}, {Hernquist}, {Pakmor}, {Kannan}, {Hern{\'a}ndez-Aguayo},
  {Barrera}, {Ferlito}, {Delgado}, {Bose}, \& {Frenk}}]{MTNG_8}
{Contreras}, S., {Angulo}, R.~E., {Springel}, V., {et~al.} 2023, \mnras, 524,
  2489

\bibitem[{{Delgado} {et~al.}(2023){Delgado}, {Hadzhiyska}, {Bose}, {Springel},
  {Hernquist}, {Barrera}, {Pakmor}, {Ferlito}, {Kannan},
  {Hern{\'a}ndez-Aguayo}, {White}, \& {Frenk}}]{MTNG_9}
{Delgado}, A.~M., {Hadzhiyska}, B., {Bose}, S., {et~al.} 2023, \mnras, 523,
  5899

\bibitem[{{Diemer}(2018)}]{Diemer_2018}
{Diemer}, B. 2018, \apjs, 239, 35

\bibitem[{{Diemer}(2020)}]{Diemer_2020}
---. 2020, \apjs, 251, 17

\bibitem[{{Diemer} \& {Kravtsov}(2014)}]{Diemer_2014}
{Diemer}, B., \& {Kravtsov}, A.~V. 2014, \apj, 789, 1

\bibitem[{{Diemer} {et~al.}(2017){Diemer}, {Sparre}, {Abramson}, \&
  {Torrey}}]{diemer_17_sfh}
{Diemer}, B., {Sparre}, M., {Abramson}, L.~E., \& {Torrey}, P. 2017, \apj, 839,
  26

\bibitem[{{Diemer} {et~al.}(2018){Diemer}, {Stevens}, {Forbes}, {Marinacci},
  {Hernquist}, {Lagos}, {Sternberg}, {Pillepich}, {Nelson}, {Popping},
  {Villaescusa- Navarro}, {Torrey}, \& {Vogelsberger}}]{diemer_18_hih2}
{Diemer}, B., {Stevens}, A. R.~H., {Forbes}, J.~C., {et~al.} 2018, The
  Astrophysical Journal Supplement Series, 238, 33

\bibitem[{{Eisenstein} \& {Hu}(1998)}]{Eisenstein_1998}
{Eisenstein}, D.~J., \& {Hu}, W. 1998, \apj, 496, 605

\bibitem[{{Farid} {et~al.}(2023){Farid}, {Aung}, {Nagai}, {Farahi}, \&
  {Rozo}}]{Farid2023}
{Farid}, D., {Aung}, H., {Nagai}, D., {Farahi}, A., \& {Rozo}, E. 2023,
  Astronomy and Computing, 45, 100743

\bibitem[{{Ferlito} {et~al.}(2023){Ferlito}, {Springel}, {Davies},
  {Hern{\'a}ndez-Aguayo}, {Pakmor}, {Barrera}, {White}, {Delgado},
  {Hadzhiyska}, {Hernquist}, {Kannan}, {Bose}, \& {Frenk}}]{MTNG_10}
{Ferlito}, F., {Springel}, V., {Davies}, C.~T., {et~al.} 2023, \mnras, 524,
  5591

\bibitem[{{Foster} {et~al.}(2012){Foster}, {Ji}, {Smith}, \&
  {Brickhouse}}]{Foster2012}
{Foster}, A.~R., {Ji}, L., {Smith}, R.~K., \& {Brickhouse}, N.~S. 2012, \apj,
  756, 128

\bibitem[{{Fournier} {et~al.}(2024){Fournier}, {Grete}, {Br{\"u}ggen},
  {Glines}, \& {O'Shea}}]{Fournier_2024}
{Fournier}, M., {Grete}, P., {Br{\"u}ggen}, M., {Glines}, F.~W., \& {O'Shea},
  B.~W. 2024, arXiv e-prints, arXiv:2406.05044

\bibitem[{{Ghirardini} {et~al.}(2024){Ghirardini}, {Bulbul}, {Artis}, {Clerc},
  {Garrel}, {Grandis}, {Kluge}, {Liu}, {Bahar}, {Balzer}, {Chiu}, {Comparat},
  {Gruen}, {Kleinebreil}, {Krippendorf}, {Merloni}, {Nandra}, {Okabe},
  {Pacaud}, {Predehl}, {Ramos-Ceja}, {Reiprich}, {Sanders}, {Schrabback},
  {Seppi}, {Zelmer}, {Zhang}, {Bornemann}, {Brunner}, {Burwitz}, {Coutinho},
  {Dennerl}, {Freyberg}, {Friedrich}, {Gaida}, {Gueguen}, {Haberl}, {Kink},
  {Lamer}, {Li}, {Liu}, {Maitra}, {Meidinger}, {Mueller}, {Miyatake},
  {Miyazaki}, {Robrade}, {Schwope}, \&
  {Stewart}}]{eROSITA_Cluster_Abundances_2024}
{Ghirardini}, V., {Bulbul}, E., {Artis}, E., {et~al.} 2024, arXiv e-prints,
  arXiv:2402.08458

\bibitem[{{Gouin} {et~al.}(2021){Gouin}, {Bonnaire}, \& {Aghanim}}]{Gouin_2021}
{Gouin}, C., {Bonnaire}, T., \& {Aghanim}, N. 2021, \aap, 651, A56

\bibitem[{{Greenberg} {et~al.}(2019){Greenberg}, {Nonnenmacher}, \&
  {Macke}}]{Greenberg_2019}
{Greenberg}, D.~S., {Nonnenmacher}, M., \& {Macke}, J.~H. 2019, arXiv e-prints,
  arXiv:1905.07488

\bibitem[{{Hadzhiyska} {et~al.}(2023{\natexlab{a}}){Hadzhiyska}, {Eisenstein},
  {Hernquist}, {Pakmor}, {Bose}, {Delgado}, {Contreras}, {Kannan}, {White},
  {Springel}, {Frenk}, {Hern{\'a}ndez-Aguayo}, {Barrera}, \& {Monica}}]{MTNG_6}
{Hadzhiyska}, B., {Eisenstein}, D., {Hernquist}, L., {et~al.}
  2023{\natexlab{a}}, \mnras, 524, 2507

\bibitem[{{Hadzhiyska} {et~al.}(2023{\natexlab{b}}){Hadzhiyska}, {Hernquist},
  {Eisenstein}, {Delgado}, {Bose}, {Kannan}, {Pakmor}, {Springel}, {Contreras},
  {Barrera}, {Ferlito}, {Hern{\'a}ndez-Aguayo}, {White}, \& {Frenk}}]{MTNG_5}
{Hadzhiyska}, B., {Hernquist}, L., {Eisenstein}, D., {et~al.}
  2023{\natexlab{b}}, \mnras, 524, 2524

\bibitem[{{Henriksen} \& {Panda}(2024)}]{Henriksen_2024}
{Henriksen}, M.~J., \& {Panda}, P. 2024, \apjl, 961, L36

\bibitem[{{Hern{\'a}ndez-Aguayo} {et~al.}(2023){Hern{\'a}ndez-Aguayo},
  {Springel}, {Pakmor}, {Barrera}, {Ferlito}, {White}, {Hernquist},
  {Hadzhiyska}, {Delgado}, {Kannan}, {Bose}, \& {Frenk}}]{MTNG_1}
{Hern{\'a}ndez-Aguayo}, C., {Springel}, V., {Pakmor}, R., {et~al.} 2023,
  \mnras, 524, 2556

\bibitem[{{Hilton} {et~al.}(2021){Hilton}, {Sif{\'o}n}, {Naess},
  {Madhavacheril}, {Oguri}, {Rozo}, {Rykoff}, {Abbott}, {Adhikari}, {Aguena},
  {Aiola}, {Allam}, {Amodeo}, {Amon}, {Annis}, {Ansarinejad}, {Aros-Bunster},
  {Austermann}, {Avila}, {Bacon}, {Battaglia}, {Beall}, {Becker}, {Bernstein},
  {Bertin}, {Bhandarkar}, {Bhargava}, {Bond}, {Brooks}, {Burke}, {Calabrese},
  {Carrasco Kind}, {Carretero}, {Choi}, {Choi}, {Conselice}, {da Costa},
  {Costanzi}, {Crichton}, {Crowley}, {D{\"u}nner}, {Denison}, {Devlin},
  {Dicker}, {Diehl}, {Dietrich}, {Doel}, {Duff}, {Duivenvoorden}, {Dunkley},
  {Everett}, {Ferraro}, {Ferrero}, {Fert{\'e}}, {Flaugher}, {Frieman},
  {Gallardo}, {Garc{\'\i}a-Bellido}, {Gaztanaga}, {Gerdes}, {Giles}, {Golec},
  {Gralla}, {Grandis}, {Gruen}, {Gruendl}, {Gschwend}, {Gutierrez}, {Han},
  {Hartley}, {Hasselfield}, {Hill}, {Hilton}, {Hincks}, {Hinton}, {Ho},
  {Honscheid}, {Hoyle}, {Hubmayr}, {Huffenberger}, {Hughes}, {Jaelani}, {Jain},
  {James}, {Jeltema}, {Kent}, {Knowles}, {Koopman}, {Kuehn}, {Lahav}, {Lima},
  {Lin}, {Lokken}, {Loubser}, {MacCrann}, {Maia}, {Marriage}, {Martin},
  {McMahon}, {Melchior}, {Menanteau}, {Miquel}, {Miyatake}, {Moodley},
  {Morgan}, {Mroczkowski}, {Nati}, {Newburgh}, {Niemack}, {Nishizawa},
  {Ogando}, {Orlowski-Scherer}, {Page}, {Palmese}, {Partridge},
  {Paz-Chinch{\'o}n}, {Phakathi}, {Plazas}, {Robertson}, {Romer}, {Carnero
  Rosell}, {Salatino}, {Sanchez}, {Schaan}, {Schillaci}, {Sehgal}, {Serrano},
  {Shin}, {Simon}, {Smith}, {Soares-Santos}, {Spergel}, {Staggs}, {Storer},
  {Suchyta}, {Swanson}, {Tarle}, {Thomas}, {To}, {Trac}, {Ullom}, {Vale}, {Van
  Lanen}, {Vavagiakis}, {De Vicente}, {Wilkinson}, {Wollack}, {Xu}, \&
  {Zhang}}]{Hilton_2021}
{Hilton}, M., {Sif{\'o}n}, C., {Naess}, S., {et~al.} 2021, \apjs, 253, 3

\bibitem[{{Ho} {et~al.}(2023){Ho}, {Soltis}, {Farahi}, {Nagai}, {Evrard}, \&
  {Ntampaka}}]{Ho_2023}
{Ho}, M., {Soltis}, J., {Farahi}, A., {et~al.} 2023, \mnras, 524, 3289

\bibitem[{{Jiang} \& {van den Bosch}(2016)}]{Jiang_2016}
{Jiang}, F., \& {van den Bosch}, F.~C. 2016, \mnras, 458, 2848

\bibitem[{{Kannan} {et~al.}(2023){Kannan}, {Springel}, {Hernquist}, {Pakmor},
  {Delgado}, {Hadzhiyska}, {Hern{\'a}ndez-Aguayo}, {Barrera}, {Ferlito},
  {Bose}, {White}, {Frenk}, {Smith}, \& {Garaldi}}]{MTNG_4}
{Kannan}, R., {Springel}, V., {Hernquist}, L., {et~al.} 2023, \mnras, 524, 2594

\bibitem[{{Kravtsov} \& {Borgani}(2012)}]{Kravtsov_2012}
{Kravtsov}, A.~V., \& {Borgani}, S. 2012, \araa, 50, 353

\bibitem[{{Krippendorf} {et~al.}(2024){Krippendorf}, {Baron Perez}, {Bulbul},
  {Kara}, {Seppi}, {Comparat}, {Artis}, {Bahar}, {Garrel}, {Ghirardini},
  {Kluge}, {Liu}, {Ramos-Ceja}, {Sanders}, {Zhang}, {Br{\"u}ggen}, {Grandis},
  \& {Weller}}]{Krippendorf_2024}
{Krippendorf}, S., {Baron Perez}, N., {Bulbul}, E., {et~al.} 2024, \aap, 682,
  A132

\bibitem[{{Lau} {et~al.}(2021){Lau}, {Hearin}, {Nagai}, \&
  {Cappelluti}}]{Lau_2021}
{Lau}, E.~T., {Hearin}, A.~P., {Nagai}, D., \& {Cappelluti}, N. 2021, \mnras,
  500, 1029

\bibitem[{{Lau} {et~al.}(2015){Lau}, {Nagai}, {Avestruz}, {Nelson}, \&
  {Vikhlinin}}]{Lau_2015}
{Lau}, E.~T., {Nagai}, D., {Avestruz}, C., {Nelson}, K., \& {Vikhlinin}, A.
  2015, \apj, 806, 68

\bibitem[{{LeCun} {et~al.}(2015){LeCun}, {Bengio}, \& {Hinton}}]{LeCun_2015}
{LeCun}, Y., {Bengio}, Y., \& {Hinton}, G. 2015, \nat, 521, 436

\bibitem[{{Lee} {et~al.}(2023){Lee}, {Cha}, {Jee}, {Nagai}, {King}, {ZuHone},
  {Chadayammuri}, {Felix}, \& {Finner}}]{Lee_2023}
{Lee}, W., {Cha}, S., {Jee}, M.~J., {et~al.} 2023, \apj, 945, 71

\bibitem[{{Mendoza} {et~al.}(2023){Mendoza}, {Mansfield}, {Wang}, \&
  {Avestruz}}]{Mendoza_2023}
{Mendoza}, I., {Mansfield}, P., {Wang}, K., \& {Avestruz}, C. 2023, \mnras,
  523, 6386

\bibitem[{{Morrison} \& {McCammon}(1983)}]{wabs}
{Morrison}, R., \& {McCammon}, D. 1983, \apj, 270, 119

\bibitem[{{Nelson} {et~al.}(2018){Nelson}, {Pillepich}, {Springel},
  {Weinberger}, {Hernquist}, {Pakmor}, {Genel}, {Torrey}, {Vogelsberger},
  {Kauffmann}, {Marinacci}, \& {Naiman}}]{Nelson_2018}
{Nelson}, D., {Pillepich}, A., {Springel}, V., {et~al.} 2018, \mnras, 475, 624

\bibitem[{{Nelson} {et~al.}(2014{\natexlab{a}}){Nelson}, {Lau}, \&
  {Nagai}}]{Nelson2014a}
{Nelson}, K., {Lau}, E.~T., \& {Nagai}, D. 2014{\natexlab{a}}, \apj, 792, 25

\bibitem[{{Nelson} {et~al.}(2014{\natexlab{b}}){Nelson}, {Lau}, {Nagai},
  {Rudd}, \& {Yu}}]{Nelson2014b}
{Nelson}, K., {Lau}, E.~T., {Nagai}, D., {Rudd}, D.~H., \& {Yu}, L.
  2014{\natexlab{b}}, \apj, 782, 107

\bibitem[{{Nelson} {et~al.}(2012){Nelson}, {Rudd}, {Shaw}, \&
  {Nagai}}]{Nelson2012}
{Nelson}, K., {Rudd}, D.~H., {Shaw}, L., \& {Nagai}, D. 2012, \apj, 751, 121

\bibitem[{{Ntampaka} {et~al.}(2019){Ntampaka}, {ZuHone}, {Eisenstein}, {Nagai},
  {Vikhlinin}, {Hernquist}, {Marinacci}, {Nelson}, {Pakmor}, {Pillepich},
  {Torrey}, \& {Vogelsberger}}]{Ntampaka2019}
{Ntampaka}, M., {ZuHone}, J., {Eisenstein}, D., {et~al.} 2019, \apj, 876, 82

\bibitem[{{Pakmor} {et~al.}(2023){Pakmor}, {Springel}, {Coles}, {Guillet},
  {Pfrommer}, {Bose}, {Barrera}, {Delgado}, {Ferlito}, {Frenk}, {Hadzhiyska},
  {Hern{\'a}ndez-Aguayo}, {Hernquist}, {Kannan}, \& {White}}]{MTNG_2}
{Pakmor}, R., {Springel}, V., {Coles}, J.~P., {et~al.} 2023, \mnras, 524, 2539

\bibitem[{{Papamakarios} {et~al.}(2019){Papamakarios}, {Nalisnick}, {Jimenez
  Rezende}, {Mohamed}, \& {Lakshminarayanan}}]{Papamakarios_2019}
{Papamakarios}, G., {Nalisnick}, E., {Jimenez Rezende}, D., {Mohamed}, S., \&
  {Lakshminarayanan}, B. 2019, arXiv e-prints, arXiv:1912.02762

\bibitem[{{Papamakarios} {et~al.}(2018){Papamakarios}, {Sterratt}, \&
  {Murray}}]{Papamakarios_2018}
{Papamakarios}, G., {Sterratt}, D.~C., \& {Murray}, I. 2018, arXiv e-prints,
  arXiv:1805.07226

\bibitem[{{Pearson}(1895)}]{Pearson_1895}
{Pearson}, K. 1895, Proceedings of the Royal Society of London Series I, 58,
  240

\bibitem[{{Pillepich} {et~al.}(2018){Pillepich}, {Springel}, {Nelson}, {Genel},
  {Naiman}, {Pakmor}, {Hernquist}, {Torrey}, {Vogelsberger}, {Weinberger}, \&
  {Marinacci}}]{Pillepich_2018}
{Pillepich}, A., {Springel}, V., {Nelson}, D., {et~al.} 2018, \mnras, 473, 4077

\bibitem[{{Pizzardo} {et~al.}(2023){Pizzardo}, {Geller}, {Kenyon}, {Damjanov},
  \& {Diaferio}}]{Pizzardo_2023}
{Pizzardo}, M., {Geller}, M.~J., {Kenyon}, S.~J., {Damjanov}, I., \&
  {Diaferio}, A. 2023, \aap, 680, A48

\bibitem[{{Planck Collaboration} {et~al.}(2016){Planck Collaboration}, {Ade},
  {Aghanim}, {Arnaud}, {Ashdown}, {Aumont}, {Baccigalupi}, {Banday},
  {Barreiro}, {Bartlett}, {Bartolo}, {Battaner}, {Battye}, {Benabed},
  {Beno{\^\i}t}, {Benoit-L{\'e}vy}, {Bernard}, {Bersanelli}, {Bielewicz},
  {Bock}, {Bonaldi}, {Bonavera}, {Bond}, {Borrill}, {Bouchet}, {Boulanger},
  {Bucher}, {Burigana}, {Butler}, {Calabrese}, {Cardoso}, {Catalano},
  {Challinor}, {Chamballu}, {Chary}, {Chiang}, {Chluba}, {Christensen},
  {Church}, {Clements}, {Colombi}, {Colombo}, {Combet}, {Coulais}, {Crill},
  {Curto}, {Cuttaia}, {Danese}, {Davies}, {Davis}, {de Bernardis}, {de Rosa},
  {de Zotti}, {Delabrouille}, {D{\'e}sert}, {Di Valentino}, {Dickinson},
  {Diego}, {Dolag}, {Dole}, {Donzelli}, {Dor{\'e}}, {Douspis}, {Ducout},
  {Dunkley}, {Dupac}, {Efstathiou}, {Elsner}, {En{\ss}lin}, {Eriksen},
  {Farhang}, {Fergusson}, {Finelli}, {Forni}, {Frailis}, {Fraisse},
  {Franceschi}, {Frejsel}, {Galeotta}, {Galli}, {Ganga}, {Gauthier}, {Gerbino},
  {Ghosh}, {Giard}, {Giraud-H{\'e}raud}, {Giusarma}, {Gjerl{\o}w},
  {Gonz{\'a}lez-Nuevo}, {G{\'o}rski}, {Gratton}, {Gregorio}, {Gruppuso},
  {Gudmundsson}, {Hamann}, {Hansen}, {Hanson}, {Harrison}, {Helou},
  {Henrot-Versill{\'e}}, {Hern{\'a}ndez-Monteagudo}, {Herranz}, {Hildebrandt},
  {Hivon}, {Hobson}, {Holmes}, {Hornstrup}, {Hovest}, {Huang}, {Huffenberger},
  {Hurier}, {Jaffe}, {Jaffe}, {Jones}, {Juvela}, {Keih{\"a}nen}, {Keskitalo},
  {Kisner}, {Kneissl}, {Knoche}, {Knox}, {Kunz}, {Kurki-Suonio}, {Lagache},
  {L{\"a}hteenm{\"a}ki}, {Lamarre}, {Lasenby}, {Lattanzi}, {Lawrence}, {Leahy},
  {Leonardi}, {Lesgourgues}, {Levrier}, {Lewis}, {Liguori}, {Lilje},
  {Linden-V{\o}rnle}, {L{\'o}pez-Caniego}, {Lubin}, {Mac{\'\i}as-P{\'e}rez},
  {Maggio}, {Maino}, {Mandolesi}, {Mangilli}, {Marchini}, {Maris}, {Martin},
  {Martinelli}, {Mart{\'\i}nez-Gonz{\'a}lez}, {Masi}, {Matarrese}, {McGehee},
  {Meinhold}, {Melchiorri}, {Melin}, {Mendes}, {Mennella}, {Migliaccio},
  {Millea}, {Mitra}, {Miville-Desch{\^e}nes}, {Moneti}, {Montier}, {Morgante},
  {Mortlock}, {Moss}, {Munshi}, {Murphy}, {Naselsky}, {Nati}, {Natoli},
  {Netterfield}, {N{\o}rgaard-Nielsen}, {Noviello}, {Novikov}, {Novikov},
  {Oxborrow}, {Paci}, {Pagano}, {Pajot}, {Paladini}, {Paoletti}, {Partridge},
  {Pasian}, {Patanchon}, {Pearson}, {Perdereau}, {Perotto}, {Perrotta},
  {Pettorino}, {Piacentini}, {Piat}, {Pierpaoli}, {Pietrobon}, {Plaszczynski},
  {Pointecouteau}, {Polenta}, {Popa}, {Pratt}, {Pr{\'e}zeau}, {Prunet},
  {Puget}, {Rachen}, {Reach}, {Rebolo}, {Reinecke}, {Remazeilles}, {Renault},
  {Renzi}, {Ristorcelli}, {Rocha}, {Rosset}, {Rossetti}, {Roudier},
  {Rouill{\'e} d'Orfeuil}, {Rowan-Robinson}, {Rubi{\~n}o-Mart{\'\i}n},
  {Rusholme}, {Said}, {Salvatelli}, {Salvati}, {Sandri}, {Santos},
  {Savelainen}, {Savini}, {Scott}, {Seiffert}, {Serra}, {Shellard}, {Spencer},
  {Spinelli}, {Stolyarov}, {Stompor}, {Sudiwala}, {Sunyaev}, {Sutton},
  {Suur-Uski}, {Sygnet}, {Tauber}, {Terenzi}, {Toffolatti}, {Tomasi},
  {Tristram}, {Trombetti}, {Tucci}, {Tuovinen}, {T{\"u}rler}, {Umana},
  {Valenziano}, {Valiviita}, {Van Tent}, {Vielva}, {Villa}, {Wade}, {Wandelt},
  {Wehus}, {White}, {White}, {Wilkinson}, {Yvon}, {Zacchei}, \&
  {Zonca}}]{Planck_2015_Values}
{Planck Collaboration}, {Ade}, P.~A.~R., {Aghanim}, N., {et~al.} 2016, \aap,
  594, A13

\bibitem[{{Pratt} {et~al.}(2019){Pratt}, {Arnaud}, {Biviano}, {Eckert},
  {Ettori}, {Nagai}, {Okabe}, \& {Reiprich}}]{Pratt_2019}
{Pratt}, G.~W., {Arnaud}, M., {Biviano}, A., {et~al.} 2019, \ssr, 215, 25

\bibitem[{{Qiu} {et~al.}(2024){Qiu}, {Napolitano}, {Borgani}, {Zhong}, {Li},
  {Radovich}, {Lin}, {Dolag}, {Tortora}, {Wang}, {Remus}, {Wu}, \&
  {Longo}}]{Qiu_2024}
{Qiu}, L., {Napolitano}, N.~R., {Borgani}, S., {et~al.} 2024, \aap, 687, A1

\bibitem[{{Rothschild} {et~al.}(2022){Rothschild}, {Nagai}, {Aung}, {Green},
  {Ntampaka}, \& {ZuHone}}]{Rothschild_2022}
{Rothschild}, T., {Nagai}, D., {Aung}, H., {et~al.} 2022, \mnras, 513, 333

\bibitem[{{Shi} {et~al.}(2016){Shi}, {Komatsu}, {Nagai}, \& {Lau}}]{Shi2016}
{Shi}, X., {Komatsu}, E., {Nagai}, D., \& {Lau}, E.~T. 2016, \mnras, 455, 2936

\bibitem[{{Shin} \& {Diemer}(2023)}]{Shin_2023}
{Shin}, T.-h., \& {Diemer}, B. 2023, \mnras, 521, 5570

\bibitem[{{Soltis} \& {Garrison}(2024)}]{Soltis_2024}
{Soltis}, J., \& {Garrison}, L.~H. 2024, \mnras, 532, 1729

\bibitem[{{Soltis} {et~al.}(2022){Soltis}, {Ntampaka}, {Wu}, {ZuHone},
  {Evrard}, {Farahi}, {Ho}, \& {Nagai}}]{Soltis_2022}
{Soltis}, J., {Ntampaka}, M., {Wu}, J.~F., {et~al.} 2022, \apj, 940, 60

\bibitem[{Springel {et~al.}(2021)Springel, Pakmor, Zier, \&
  Reinecke}]{Springel_2021}
Springel, V., Pakmor, R., Zier, O., \& Reinecke, M. 2021, Monthly Notices of
  the Royal Astronomical Society, 506, 2871

\bibitem[{{Springel} {et~al.}(2005){Springel}, {White}, {Jenkins}, {Frenk},
  {Yoshida}, {Gao}, {Navarro}, {Thacker}, {Croton}, {Helly}, {Peacock}, {Cole},
  {Thomas}, {Couchman}, {Evrard}, {Colberg}, \& {Pearce}}]{Springel_2005}
{Springel}, V., {White}, S. D.~M., {Jenkins}, A., {et~al.} 2005, \nat, 435, 629

\bibitem[{{Srisawat} {et~al.}(2013){Srisawat}, {Knebe}, {Pearce}, {Schneider},
  {Thomas}, {Behroozi}, {Dolag}, {Elahi}, {Han}, {Helly}, {Jing}, {Jung},
  {Lee}, {Mao}, {Onions}, {Rodriguez-Gomez}, {Tweed}, \&
  {Yi}}]{halo_finder_comp}
{Srisawat}, C., {Knebe}, A., {Pearce}, F.~R., {et~al.} 2013, \mnras, 436, 150

\bibitem[{{Sunyaev} \& {Zeldovich}(1972)}]{SZ_1972}
{Sunyaev}, R.~A., \& {Zeldovich}, Y.~B. 1972, Comments on Astrophysics and
  Space Physics, 4, 173

\bibitem[{{Sweere} {et~al.}(2022){Sweere}, {Valtchanov}, {Lieu}, {Vojtekova},
  {Verdugo}, {Santos-Lleo}, {Pacaud}, {Briassouli}, \& {C{\'a}mpora
  P{\'e}rez}}]{Sweere_2022}
{Sweere}, S.~F., {Valtchanov}, I., {Lieu}, M., {et~al.} 2022, \mnras, 517, 4054

\bibitem[{{Vall{\'e}s-P{\'e}rez} {et~al.}(2023){Vall{\'e}s-P{\'e}rez},
  {Planelles}, {Monllor-Berbegal}, \& {Quilis}}]{Perez_2023}
{Vall{\'e}s-P{\'e}rez}, D., {Planelles}, S., {Monllor-Berbegal}, {\'O}., \&
  {Quilis}, V. 2023, \mnras, 519, 6111

\bibitem[{{Vall{\'e}s-P{\'e}rez} {et~al.}(2020){Vall{\'e}s-P{\'e}rez},
  {Planelles}, \& {Quilis}}]{Valles_Perez_2020}
{Vall{\'e}s-P{\'e}rez}, D., {Planelles}, S., \& {Quilis}, V. 2020, \mnras, 499,
  2303

\bibitem[{{Weaver} {et~al.}(2023){Weaver}, {Aung}, {Cornwell}, {Nagai}, \&
  {Arag{\'o}n-Salamanca}}]{Weaver_2023}
{Weaver}, T.~S., {Aung}, H., {Cornwell}, D.~J., {Nagai}, D., \&
  {Arag{\'o}n-Salamanca}, A. 2023, Research Notes of the American Astronomical
  Society, 7, 268

\bibitem[{{Wong} \& {Taylor}(2012)}]{Wong_2012}
{Wong}, A. W.~C., \& {Taylor}, J.~E. 2012, \apj, 757, 102

\bibitem[{{Xu} {et~al.}(2015){Xu}, {Wang}, {Chen}, \& {Li}}]{LeakyReLU_2015}
{Xu}, B., {Wang}, N., {Chen}, T., \& {Li}, M. 2015, arXiv e-prints,
  arXiv:1505.00853

\bibitem[{{Zhuravleva} {et~al.}(2023){Zhuravleva}, {Chen}, {Churazov},
  {Schekochihin}, {Zhang}, \& {Nagai}}]{Zhuravleva_2023}
{Zhuravleva}, I., {Chen}, M.~C., {Churazov}, E., {et~al.} 2023, \mnras, 520,
  5157

\bibitem[{{ZuHone} {et~al.}(2014){ZuHone}, {Biffi}, {Hallman}, {Randall},
  {Foster}, \& {Schmid}}]{ZuHone2014}
{ZuHone}, J.~A., {Biffi}, V., {Hallman}, E.~J., {et~al.} 2014, arXiv e-prints,
  arXiv:1407.1783

\end{thebibliography}

\end{document}